\def\cm{{\rm\thinspace cm}}

\def\K{{\rm\thinspace K}} \def\keV{{\rm\thinspace keV}}
\def\km{{\rm\thinspace km}} 
 
\def\Mpc{{\rm\thinspace Mpc}} 
\def\Msun{\hbox{$\rm\thinspace M_{\odot}$}}

\def\s{{\rm\thinspace s}} \def\yr{{\rm\thinspace yr}}

\def\pcmcu{\hbox{$\cm^{-3}\,$}}

\def\kmps{\hbox{$\km\s^{-1}\,$}}

\def\Msunpyr{\hbox{$\Msun\yr^{-1}\,$}} 
\def\pcmsq{\hbox{$\cm^{-2}\,$}}

\def\kmpspMpc{\hbox{$\kmps\Mpc^{-1}$}}

\documentclass[usegraphicx]{mn2e}

\usepackage{amssymb}
\usepackage{mathptmx}

\begin{document}

\title[Very deep \emph{Chandra} observation of the Perseus cluster]
{A very deep \emph{Chandra} observation of the Perseus cluster: shocks,
ripples and conduction}
\author[A.C. Fabian et al]
{\parbox[]{6.in} {A.C. Fabian$^1$\thanks{E-mail: acf@ast.cam.ac.uk},
  J.S.~Sanders$^1$, G.B.~Taylor$^{2,3,4}$, S.W.~Allen$^{1,2}$,\\
C.S.~Crawford$^1$, R.M.~Johnstone$^1$ and K.~Iwasawa$^1$\\
\footnotesize
  $^1$ Institute of Astronomy, Madingley Road, Cambridge CB3 0HA\\
  $^2$ Kavli Institute for Particle Astrophysics and Cosmology,
Stanford University, 382 Via Pueblo Mall, Stanford, CA 94305-4060, USA\\
  $^3$ National Radio Astronomy Observatory, Socorro, NM 87801, USA\\
  $^4$ University of New Mexico, Dept. of Physics \& Astronomy,
Albuquerque, NM 87131, USA
}}

\maketitle

\begin{abstract} We present the first results from a very deep
\emph{Chandra} X-ray observation of the core of the Perseus cluster of
galaxies. A pressure map reveals a clear thick band of high pressure
around the inner radio bubbles. The gas in the band must be expanding
outward and the sharp front to it is identified as a shock front, yet
we see no temperature jump across it; indeed there is more soft
emission behind the shock than in front of it. We conclude that in
this inner region either thermal conduction operates efficiently or
the co-existing relativistic plasma seen as the radio mini-halo is
mediating the shock. If common, isothermal shocks in cluster cores
mean that we cannot diagnose the expansion speed of radio bubbles from
temperature measurements alone. They can at times expand more rapidly
than currently assumed without producing significant regions of hot
gas. Bubbles may also be significantly more energetic.  The pressure
ripples found in earlier images are identified as isothermal sound
waves. A simple estimate based on their amplitude confirms that they
can be an effective distributed heat source able to balance radiative
cooling. We see multiphase gas with about $10^9\Msun$ at a temperature
of about 0.5~keV. Much, but not all, of this X-ray emitting cooler gas
is spatially associated with the optical filamentary nebula around the
central galaxy, NGC\,1275. A residual cooling flow of about
$50\Msunpyr$ may be taking place. A channel is found in the pressure
map along the path of the bubbles, with indications found of outer
bubbles. The channel connects in the S with a curious cold front.

\end{abstract}

\begin{keywords}
  X-rays: galaxies --- galaxies: clusters: individual: Perseus ---
  intergalactic medium
\end{keywords}

\section{Introduction} 
The Perseus cluster, A\,426, is the X-ray brightest cluster in the Sky
and has therefore been well studied by all X-ray telescopes. The X-ray
emission is due to thermal bremsstrahlung and line radiation from the
hot IntraCluster Medium (ICM) and is sharply peaked on the cluster
core, centred on the cD galaxy NGC\,1275. Jets from the nucleus of
that galaxy have inflated bubbles to the immediate N and S, displacing
the ICM (B\"ohringer et al 1993; Fabian et al 2000). Ghost bubbles
devoid of radio-emitting electrons, presumably from past activity, are
seen to the NW and S. The radiative cooling time of the gas in the
inner few tens of kpc is 2--3 hundred Myr, leading to a cooling flow of a
few $100\Msunpyr$ if there is no balancing heat input. Energy from the
bubbles or the bubble inflation process is a likely source of heat but
the energy transport and dissipation mechanisms have been uncertain. 

\begin{figure}
  \includegraphics[width=\columnwidth]{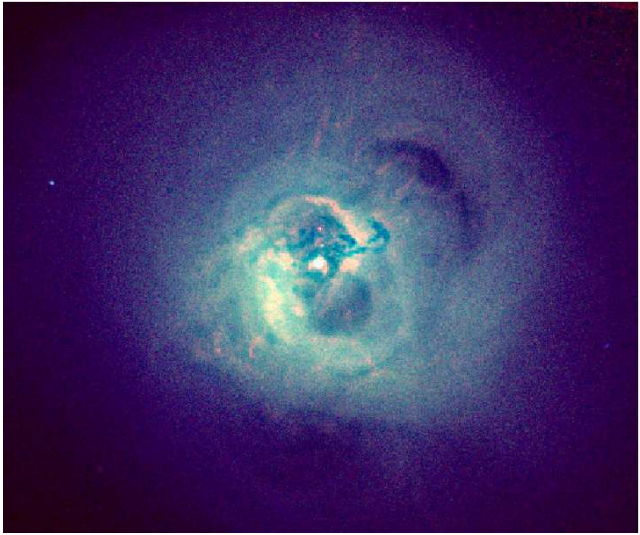}
  \caption{Colour image assembled from separate images in the 0.3--1.2
    (red), 1.2--2 (green) and 2--7 keV (blue) bands.}
\end{figure}

We have previously observed the Perseus cluster with the \emph{Chandra}
Observatory for 25~ks (Fabian et al 2000; Schmidt et al 2002; Fabian
et al 2002), 200~ks (Fabian et al 2003a,b, Sanders et al 2004, 2005)
and now present here the first results from a further 800~ks of
observation. The total good exposure time is 900~ks. 

In the earlier work we discovered both cool gas and shocks surrounding
the inner bubbles as well as quasi-circular ripples in the surrounding
gas which we interpreted as sound waves generated by the cyclical
bubbling of the central radio source. Related features have been seen
in the Virgo cluster (Forman et al 2003). The NW ghost bubble has a
horseshoe-shaped optical H$\alpha$ filament trailing it which we
interpret as showing the streamlines in the ICM. On this basis we
concluded that the ICM is not highly turbulent and thus that viscosity
is high enough to dissipate the energy carried by the sound waves
(Fabian et al 2003a,b).  Such an energy transport and dissipation
mechanism is roughly isotropic and can thereby provide the required
gently distributed heat source required by observation of this and
other similarly X-ray peaked clusters (Ruszkowski et al 2004a,b,2005;
Reynolds et al 2005; Fabian et al 2005).

Our goal in the present work is to determine the temperature and
pressure of the ICM accurately so that we can study the processes
taking place there in more detail. We indeed confirm that the pressure
jumps at the weak shock surrounding the inner bubbles and also that
the ripples represent significant ripples in pressure. The temperature
does not jump at the shock, however, which may be  due to the
action of efficient thermal conduction. The energy from the bubbles
propagates through isothermal sound waves and conduction in the inner
regions. If this is a common property of such regions then some of the
otherwise puzzling behaviour can be understood.

The redshift of the Perseus cluster is 0.0183, which for a Hubble
constant of $71\kmpspMpc$ corresponds to a luminosity distance of
78.4~Mpc and an angular scale of 367~pc per arcsec.

\section{The Data}
The \emph{Chandra} observations used for the analysis presented in
this paper are listed in Table~\ref{tab:obs}. The total exposure time
of just over 1~Ms is reduced after removing periods containing flares
to 890~ks. To filter the datasets we examined the lightcurve between
2.5 and 7~keV on the ACIS-S1 CCD. The S1 CCD is back-illuminated like
the S3 and so it is best CCD to search for flares, as the Perseus
cluster emission is dominant over flares on the S3 CCD. The
\textsc{ciao 3.3.2} \textsc{lc\_clean} tool was used to remove periods
from the dataset which deviated away from the median count rate of all
the observations.  Observations 3209 and 4289 did not use the S1 CCD
and did not show any flares on the S3 CCD, and so were left
unfiltered.

\begin{table*}
  \begin{tabular}{lllllll}
    Obs. ID & Sequence & Observation date & Exposure (ks) & Nominal roll
    (deg) & Pointing RA & Pointing Dec \\ \hline
    3209 & 800209 & 2002-08-08 &  95.8 & 101.2 & 3:19:46.86 & +41:31:51.3 \\
    4289 & 800209 & 2002-08-10 &  95.4 & 101.2 & 3:19:46.86 & +41:31:51.3 \\
    6139 & 800397 & 2004-10-04 &  51.6 & 125.9 & 3:19:45.54 & +41:31:33.9 \\
    4946 & 800397 & 2004-10-06 &  22.7 & 127.2 & 3:19:45.44 & +41:31:33.2 \\
    4948 & 800398 & 2004-10-09 & 107.5 & 128.9 & 3:19:44.75 & +41:31:40.1 \\
    4947 & 800397 & 2004-10-11 &  28.7 & 130.6 & 3:19:45.17 & +41:31:31.3 \\
    4949 & 800398 & 2004-10-12 &  28.8 & 130.9 & 3:19:44.57 & +41:31:38.7 \\
    4950 & 800399 & 2004-10-12 &  73.4 & 131.1 & 3:19:43.97 & +41:31:46.1 \\
    4952 & 800400 & 2004-10-14 & 143.2 & 132.6 & 3:19:43.22 & +41:31:52.2 \\
    4951 & 800399 & 2004-10-17 &  91.4 & 135.2 & 3:19:43.57 & +41:31:42.6 \\
    4953 & 800400 & 2004-10-18 &  29.3 & 136.2 & 3:19:42.83 & +41:31:48.5 \\
    6145 & 800397 & 2004-10-19 &  83.1 & 137.7 & 3:19:44.66 & +41:31:26.7 \\
    6146 & 800398 & 2004-10-20 &  39.2 & 138.7 & 3:19:43.92 & +41:31:32.7 \\
  \end{tabular}
  \caption{The \emph{Chandra} observations included in this
    The exposure given is the time remaining after filtering the
    lightcurve for flares. The observations were taken with the
    aimpoint on the ACIS-S3 CCD. Positions are in J2000 coordinates.}
  \label{tab:obs}
\end{table*}

The level 1 event files were reprocessed using the PSU CTI corrector
(Charge Transfer Inefficiency; Townsley et al 2002a, 2002b). Level 2
event files were produced by removing standard grades and bad time
intervals. Each of the event files was then reprojected to match the
coordinates of the 04952 observation. Images of the data in this paper
were produced by summing all the images from the individual datasets.
To correct for exposure variation we created exposure maps for each of
the CCDs for each of the datasets and for each of the bands. The
summed images were then divided by the summed exposure maps.

We have produced unsharp-mask images by subtracting images which have
been smoothed on two lengthscales. Fig.~2 (top) shows the result after
using Gaussian smoothing of 2 and 20 pixels. The ripples are very
clear, out to radii of 3--4 arcmin (60--80~kpc) from the nucleus. An
arclike step in surface brightness occurs $\sim1.5$~arcmin S of the
nucleus.  A cold front is seen to the SE (we verify that the pressure
is approximately continuous across the sharp surface brightness change
in Section 4). Such features were first seen in \emph{Chandra} images of
clusters by Markevich et al (2000). There is a major difference with
the feature seen here, however, since it is concave and cannot be due
to the core moving through a wider hotter gas. It does however appear
to connect the `bay' to the S of the nucleus which connects in towards
the nucleus along a narrow channel emerging to the SSW from the inner
regions. This corresponds spatially to a weak outer H$\alpha$ filament
(Sanders et al 2005) although extends much further than any optical
emission. X-ray emission is also associated with a much more dominant
long radial H$\alpha$ filament seen to the N of the nucleus (see e.g.
Conselice et al 2001). The X-ray feature appears to break beyond the
ripples and is labelled as the H$\alpha$ fountain.

As will be discussed later, we suspect that the radial features are
due to cold and cooler gas dragged out from the centre by rising
buoyant bubbles. They represent the main axis along which most of the
bubbles rise. The S cold front could then be the edge of a giant
hotter bubble either produced by a past major outburst of the nucleus
(cf. McNamara et al 2005 for Hydra~A) or where the hot gas
accumulates due to the interior entropy of the bubbles matching the
external value there.

\begin{figure*}
  \includegraphics[width=2\columnwidth]{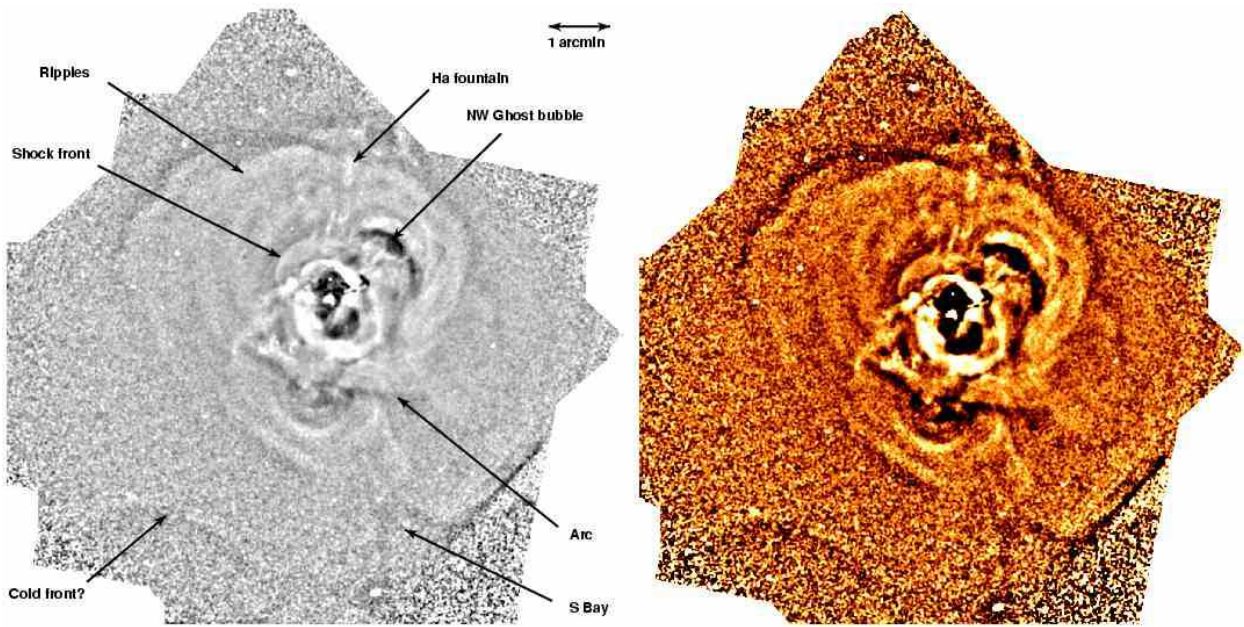}
  \caption{Unsharp mask image made from the whole 0.3-7 keV band 
  by subtracting an image smoothed with a 
  Gaussian of dispersion 10 arcsec from one smoothed by 1 arcsec and dividing 
  by the sum of the two images. Various features are
    labelled on the lower contrast image at the left.  }
\end{figure*}

\begin{figure*}
  \includegraphics[width=2\columnwidth]{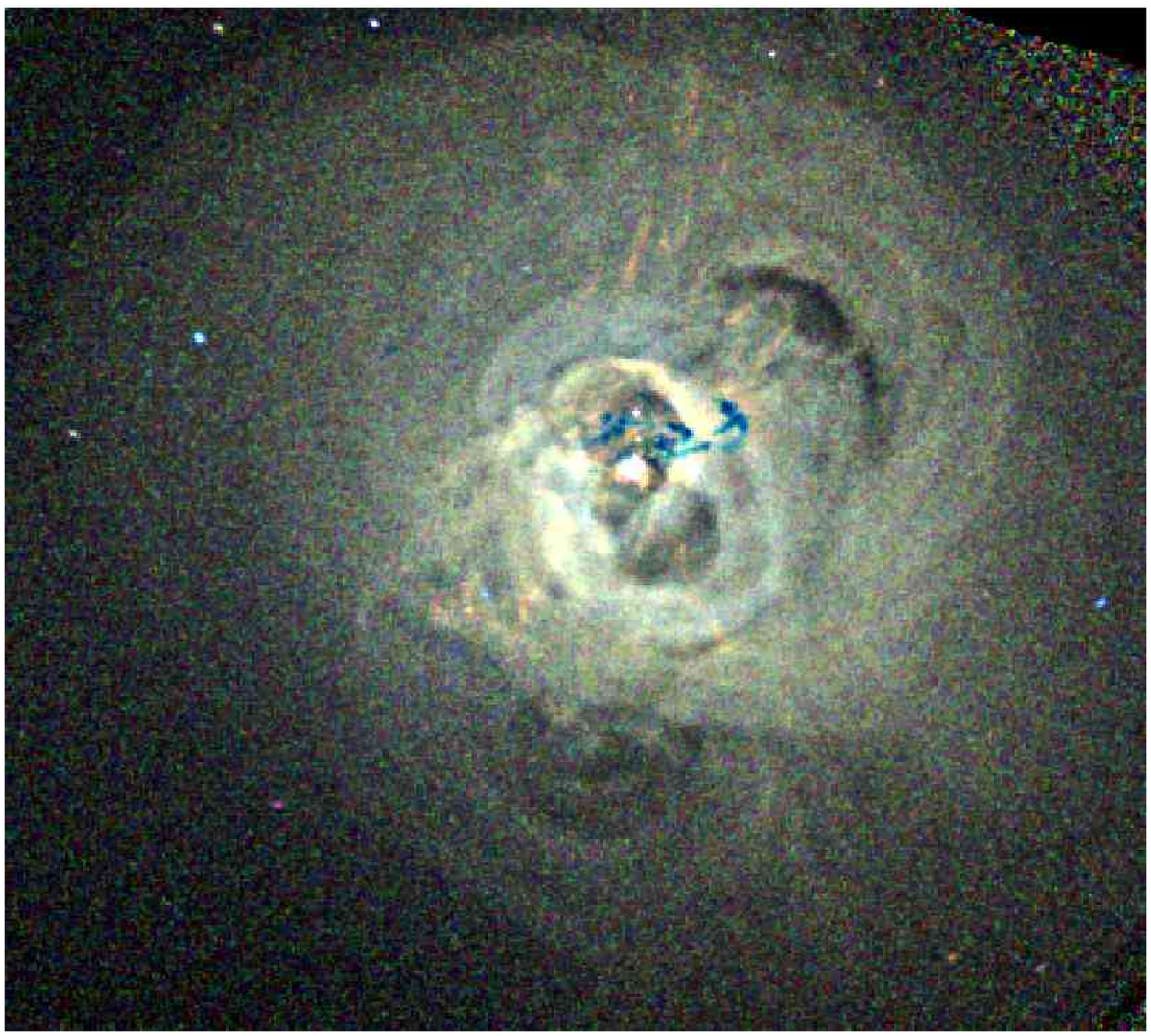}
  \caption{Colour image made from the 0.3-1.2 (red), 1.2-2. (green),
    2.-7 keV (blue) bands. A 10 arcsec smoothed image has been scaled
    to 80 per cent of its intensity and then subtracted in order to
    bring out fainter features lost in the high intensity range of raw
    images.  The blue structure to the N of the nucleus is caused by
    absorption in the infalling high velocity system, projected at
    least 60~kpc in front of the nucleus of NGC\,1275 (Gillmon et al
    2004).}
\end{figure*}

\section{Temperature and Pressure maps}
\begin{figure}
  \includegraphics[width=\columnwidth]{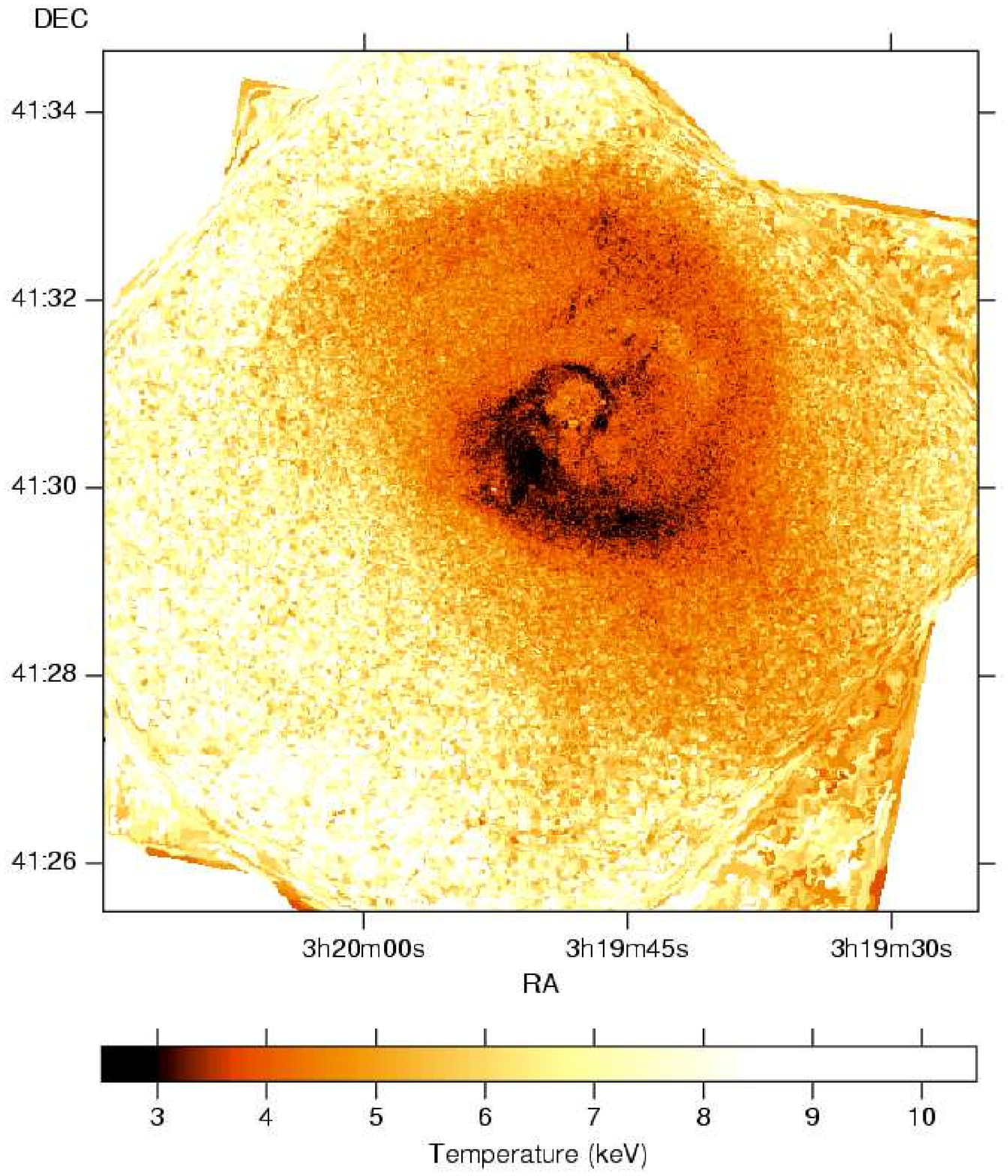}
  \caption{Temperature map calculated by fitting spectra with
    approximately 625 counts or greater. The uncertainties of the
    individual fits range from 8~per~cent in the coolest regions to
    20~per~cent in the hottest parts, ignoring the uncertainty on the
    metallicity and absorbing column density.}
  \label{fig:tmap}
\end{figure}

The total of $\sim70$ million counts in the final all band image from
the ACIS-S3 chip means that we can measure spectral properties on
unprecedented small scales.  In order to proceed we have divided the
image into bins with approximately the same number of counts and used
\textsc{xspec} 11.3.2 (Arnaud 1996) fitting with \textsc{mekal} models
(Mewe, Gronenschild \& van den Oord 1985; Liedahl, Osterheld \&
Goldstein 1995) to obtain spectral parameters, fitting between 0.5 and
7~keV. The temperature map shown in Fig.~\ref{fig:tmap} was derived in
this way using a contour binning approach (Sanders in preparation)
with 625 or greater ct per spectrum. In each fit the metallicity (in
Solar ratios; Anders \& Grevesse 1989) and absorption column density
were fixed at values measured when fitting spectra from bins
containing $10^4$ ct per bin or greater, except in the region around
the High Velocity System where the absorbing column density was
allowed to be free.  The results for these parameters is broadly
similar to those found in our earlier work (Sanders et al 2004).
Details will be given in a later paper. For the present work we
concentrate on the temperature and emission measure distributions.

We used standard blank sky observations to act as backgrounds for the
spectral fitting. The background observations were split into sections
to match the ratio of exposure time between each foreground
observation. These datasets were then reprojected to match the
foreground observations, and then reprojected to the 4952
observation. The exposure time of the backgrounds were altered to
ensure the same rate of counts between 9 and 12~keV as their
respective foregrounds, in order to correct for the variation of
background with time. To create a total spectrum, the spectra from
each of the individual observations were added together, excluding
observations which did not have any counts in the region examined. The
background spectra were added together similarly. The standard PSU CTI
corrector response was used. Ancillary responses for each dataset and
region were produced using the \textsc{ciao} \textsc{mkwarf} tool,
weighting CCD regions using the number of counts between 0.5 and
7~keV.  These ancillary responses were averaged for each region,
weighting according to the number of counts between 0.5 and 7~keV for
a particular dataset.

The temperature map (Fig.~4) shows in great detail the `swirl' around
NGC\,1275 (Churazov et al 2000). Whether the swirl is really a single
connected structure or an outer ring partially opened on the E and
connected to the rim of the inner N bubble (Dunn et al 2005b) remains
unclear. Some `fountaining' can be seen to the N of this N
bubble. This is associated with the N optical H$\alpha$ filaments
which are surrounded by gas at about $1\keV$ (see Fig.~3). A
disruption in the outer ring is seen to the SE of the nucleus
coincident with the optical `blue loop'; this is discussed in Section
6.

We now focus on measurements of entropy $S$ and particularly the
pressure $P$ of the gas.  A simple method for obtaining these
quantities is to assume that the density $n$ is proportional to the
square root of the X-ray surface brightness and then use $P=nkT$ and
$S=Tn^{-2/3}$. Here we use a slightly better approach based on the
emission measure, $A$, obtained from spectral fits.  This is
proportional to $n^2 V$ where $V$ is the volume along the line of
sight. Since the emission is strongly peaked we ignore $V$ at this
stage and produced `projected' entropy and pressure maps (Fig.~5).
 
\begin{figure*}
  \includegraphics[width=2\columnwidth]{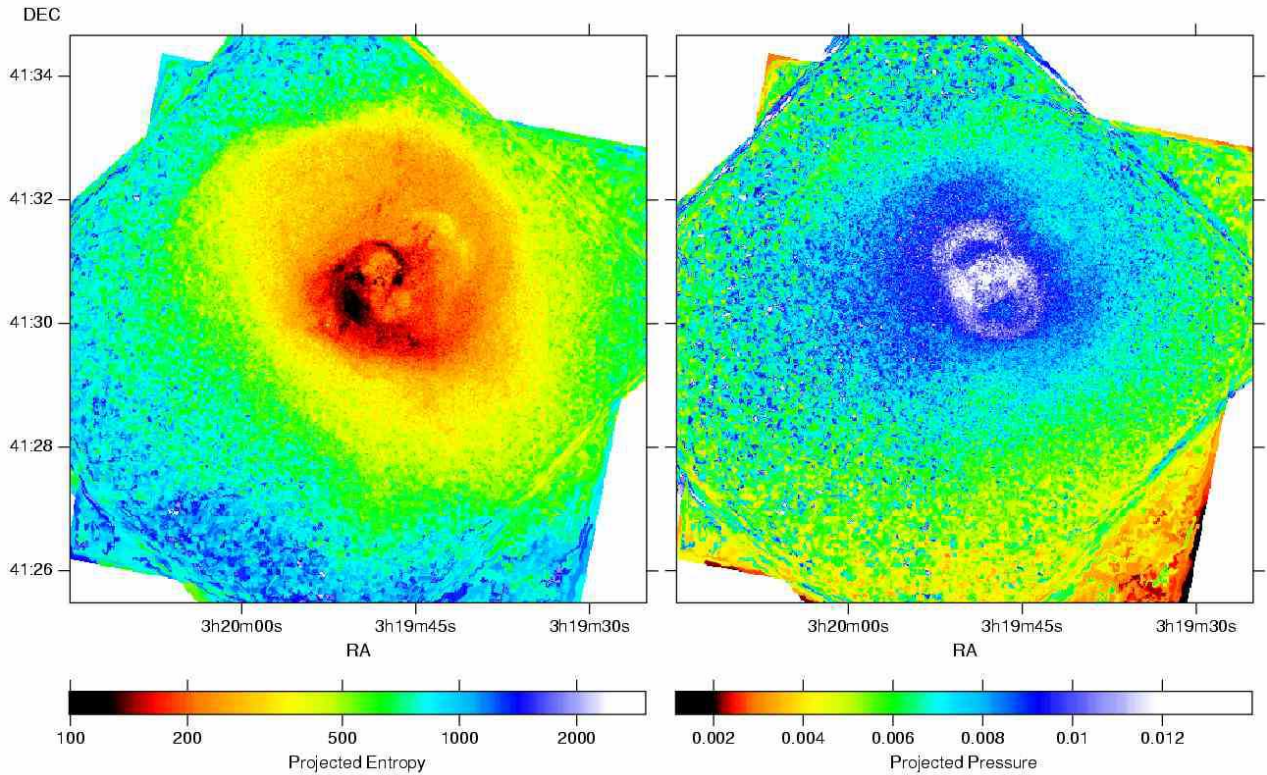}
  \caption{Entropy (left) and pressure (right) maps. The entropy map
    was calculated using $kT \: A^{-1/3}$, in units of
    keV~cm${}^{5/3}$~arcsec${}^{2/3}$, where $A$ is the \textsc{mekal}
    normalisation per square arcsecond. The pressure map was
    calculated using $kT \: A^{1/2}$, in units of
    keV~cm${}^{-5/2}$~arcsec${}^{-1}$. These maps were generated by
    fitting regions containing approximately 625 counts or greater.}
  \label{fig:entropy_pressure}
\end{figure*}

The entropy map (Fig.~\ref{fig:entropy_pressure} left) emphasises
where gas may have cooled and resembles the temperature map. The
pressure map (Fig.~\ref{fig:entropy_pressure} right) on the other hand
shows clearly a thick band around the inner radio bubbles and little
sign of azimuthal asymmetry. As found by Sanders et al (2004) the
pressure distribution is reasonably circularly symmetric, as expected
for gas close to hydrostatic equilibrium. This is not just a
consequence of our volume assumption, since we see that the `swirl' in
temperature has completely disappeared, as has the arc noted in
Fig.~2.  

A thick, higher pressure band surrounds the radio-filled cavities or
bubbles (Fig.~6). This presumably is shocked gas produced by the
inflation of the bubbles. It is remarkable that we see it as two
mostly complete rings in the projected pressure map. This means that
the two bubbles cannot lie in the plane of the Sky but must be
arranged so that one is nearer us than the other. Since the nearer
radio jet is the S one based on VLBI radio data, we suppose that the
nearer bubble is the S one.

\begin{figure}
  \includegraphics[width=\columnwidth]{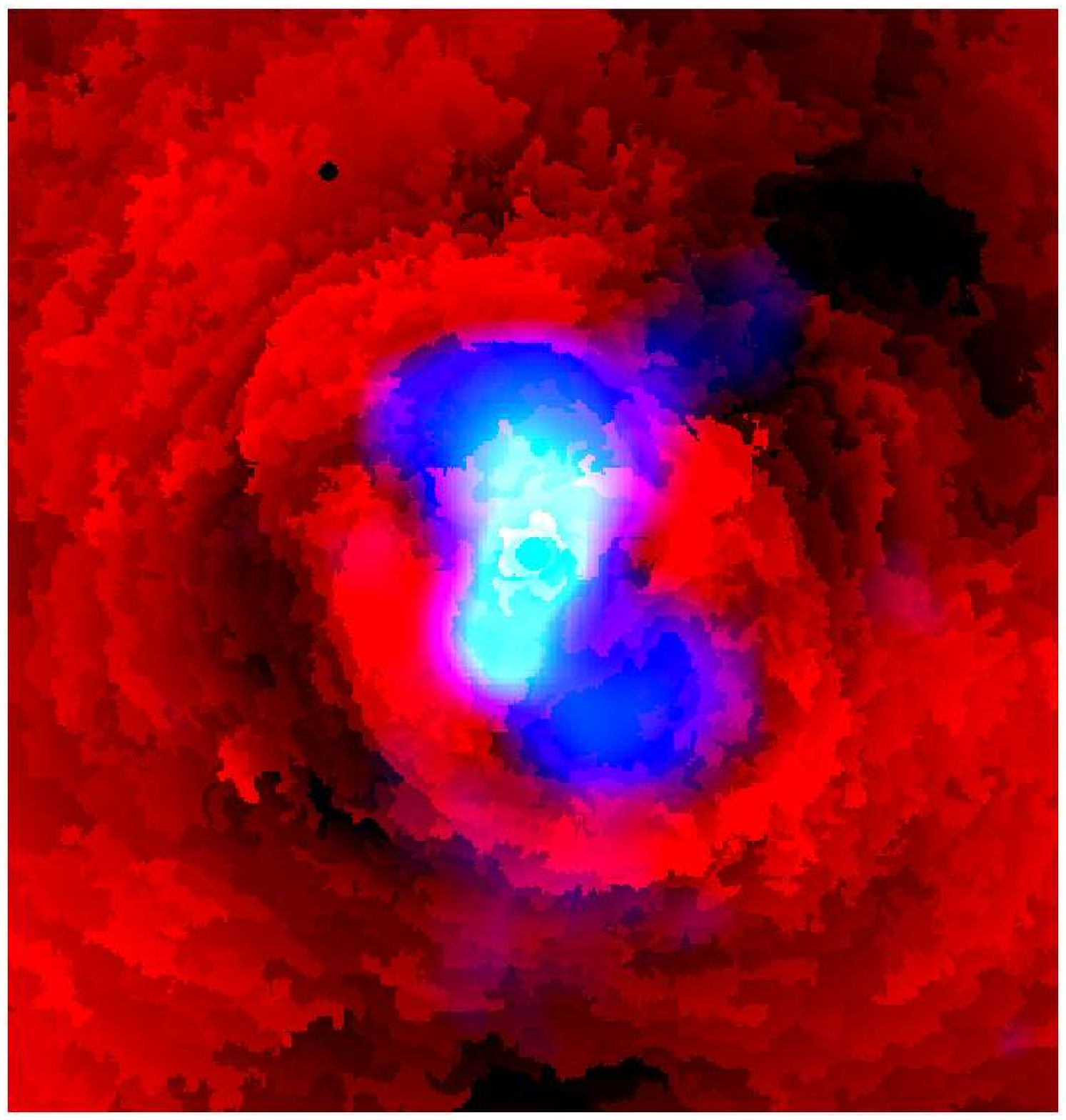}
  \caption{1.4~GHz Radio map in blue superimposed on the pressure difference
    map in red, where the average pressure at each radius has been
    subtracted. In this map the temperatures and normalisations were
    measured using regions containing approximately $10^4$ counts or
    greater.}
  \label{fig:deltaP_radio}
\end{figure}

\begin{figure} 
\includegraphics[width=\columnwidth]{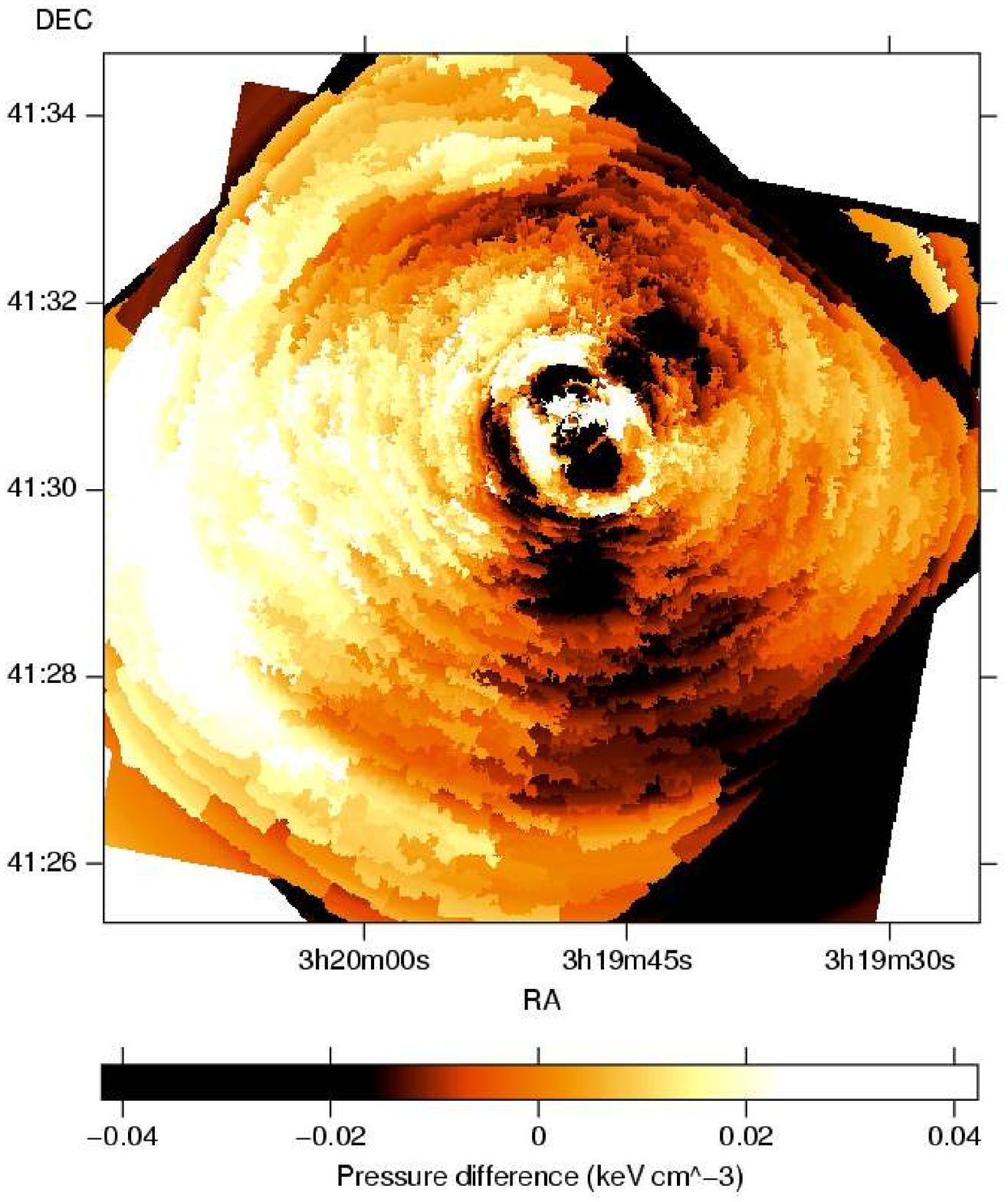}
\caption{Thermal pressure map where the mean pressure at each radius
has been subtracted. In this map the temperatures and normalisations
were measured using regions containing approximately $10^4$ counts or
greater. Note that the 'channel' caused by a sequence of 4 thermal
pressure dips running to the S of the nucleus. The outer ones are
assumed to old ghost bubbles and the missing pressure is assumed to be
due to relativistic plasma. A twisted channel is also seen to the N. } 
\label{fig:deltaP} 
\end{figure}

There is some azimuthal asymmetry in the pressure map, mostly
associated with the bubbles. In order to see this we have subtracted
the mean pressure at each radius to produce the pressure difference
map (Figs \ref{fig:deltaP_radio} and \ref{fig:deltaP}). There are
clearly some lower pressure regions to the N and S, probably
associated with older, outer bubbles. The region to the SSW has a
higher metallicity likely due to older bubbles dragging metal-rich gas
there (Sanders et al 2005).

To the south we see two further tangential arclike pressure minima
beyond the outer S bubble. These coincide with the high abundance
shell reported by Sanders et al (2005). To the N we also see a large
arclike pressure minimum. 

We suspect that these arclike pressure minima are old bubbles. The
large size of these bubbles could indicate that the activity was much
stronger in the past, so blowing larger bubbles, or may just be due to
bubbles merging. Surrounding gas may leak into the bubbles so making
them less buoyant, or magnetic fields and structures may be important.

No pressure jump is associated with the concave structure to the S,
confirming that it is part of a cold front. 

\section{The shock and ripples}

\begin{figure*}
  \includegraphics[width=.9\columnwidth]{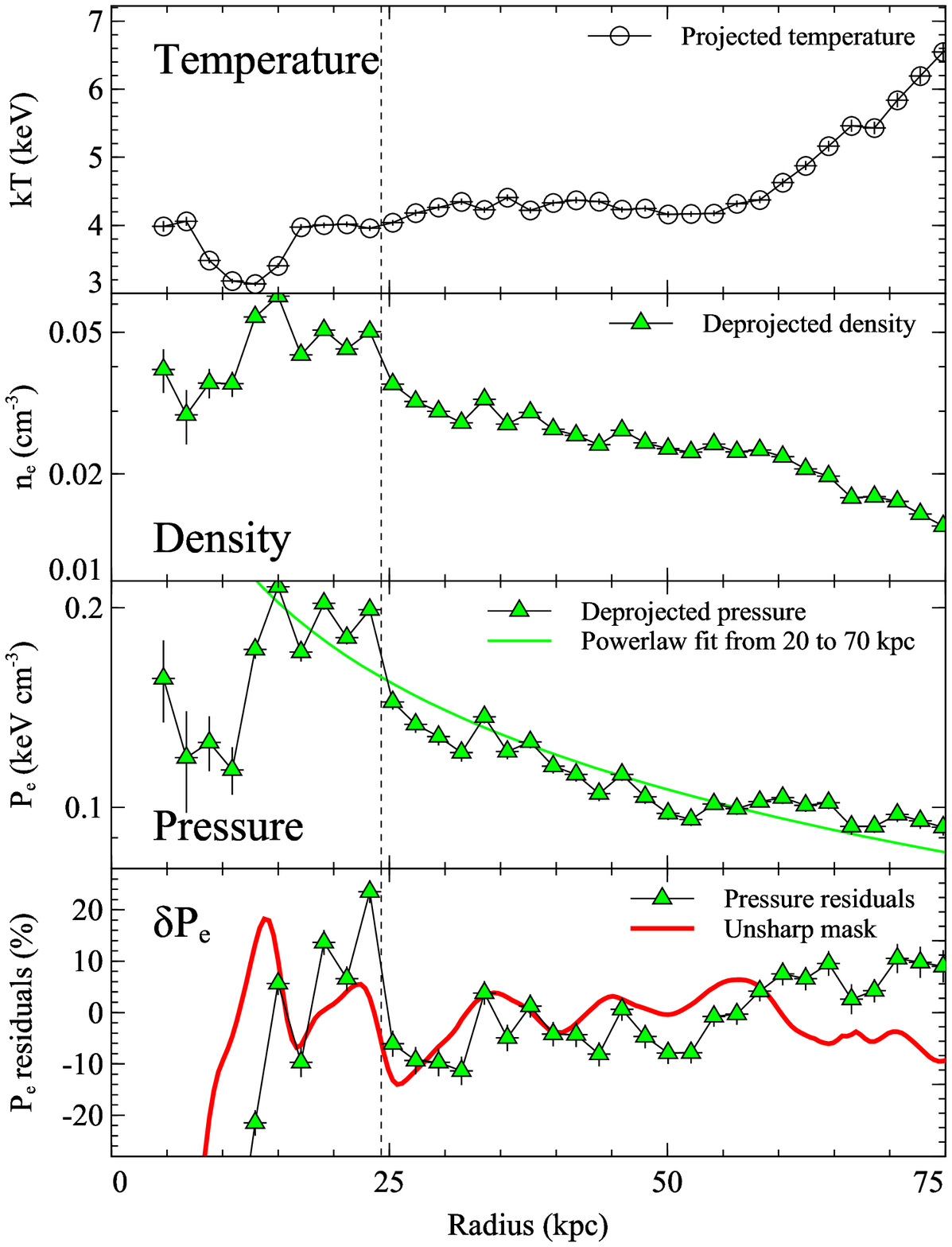}
  \includegraphics[width=.9\columnwidth]{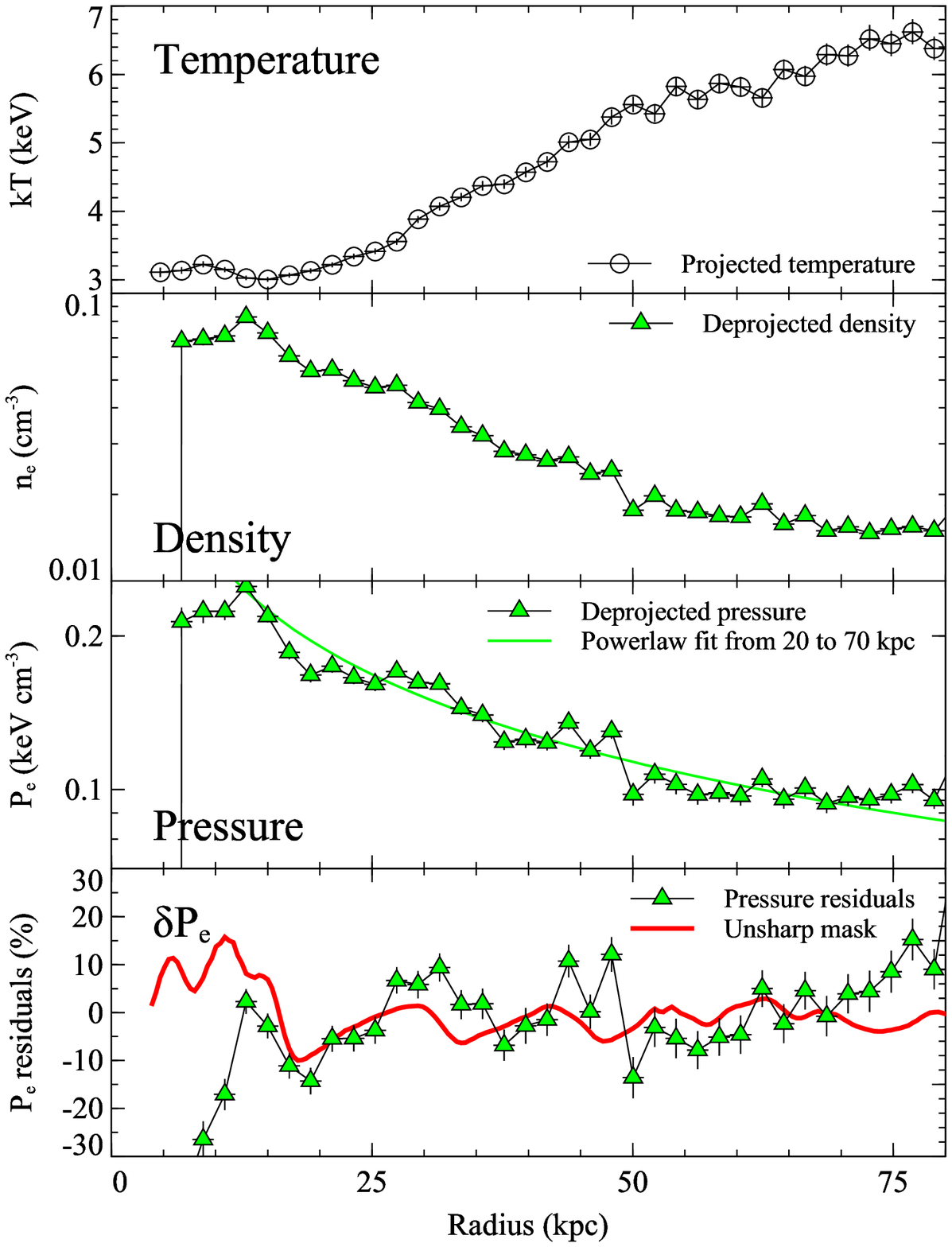}
  \includegraphics[width=.9\columnwidth]{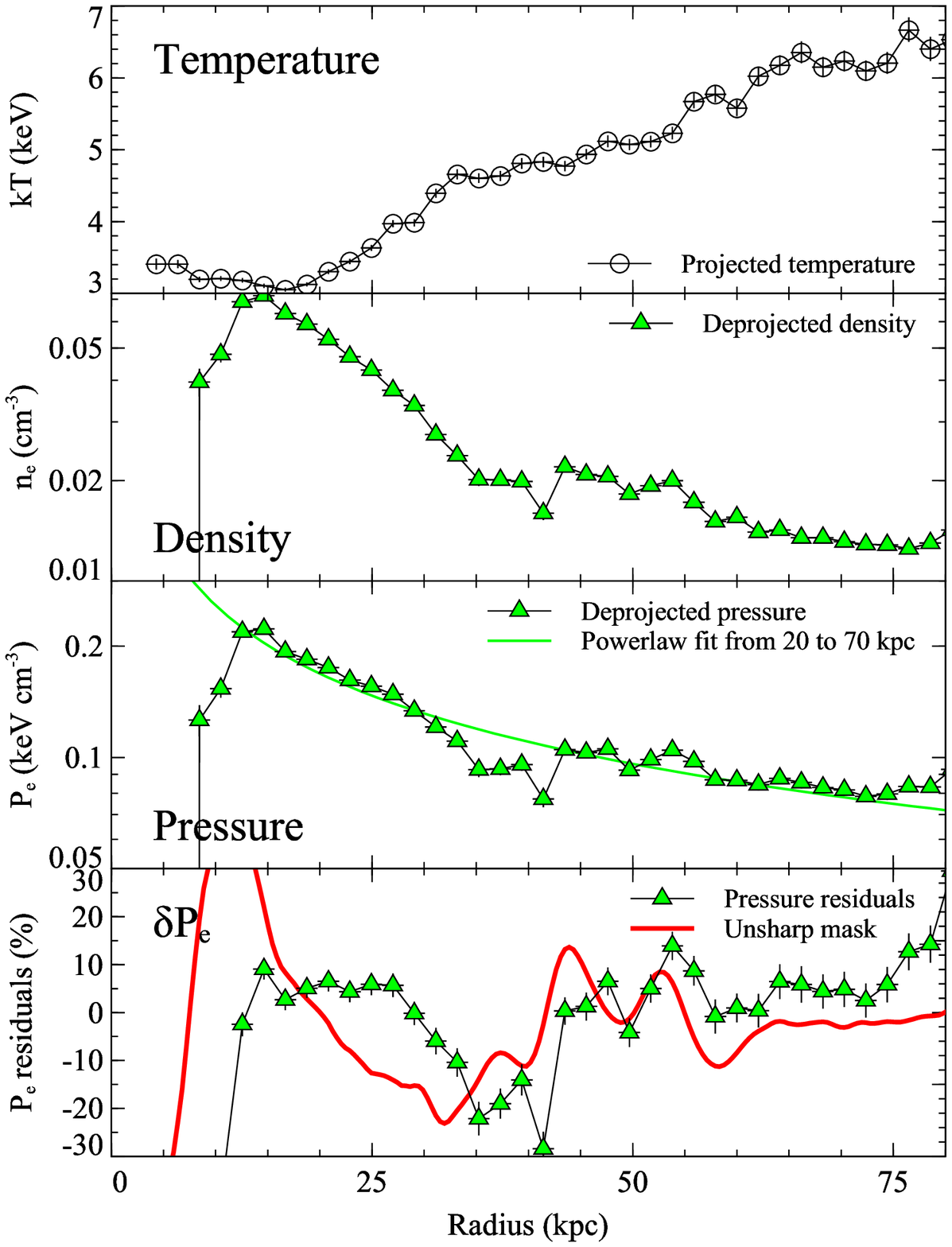}
  \includegraphics[width=.9\columnwidth]{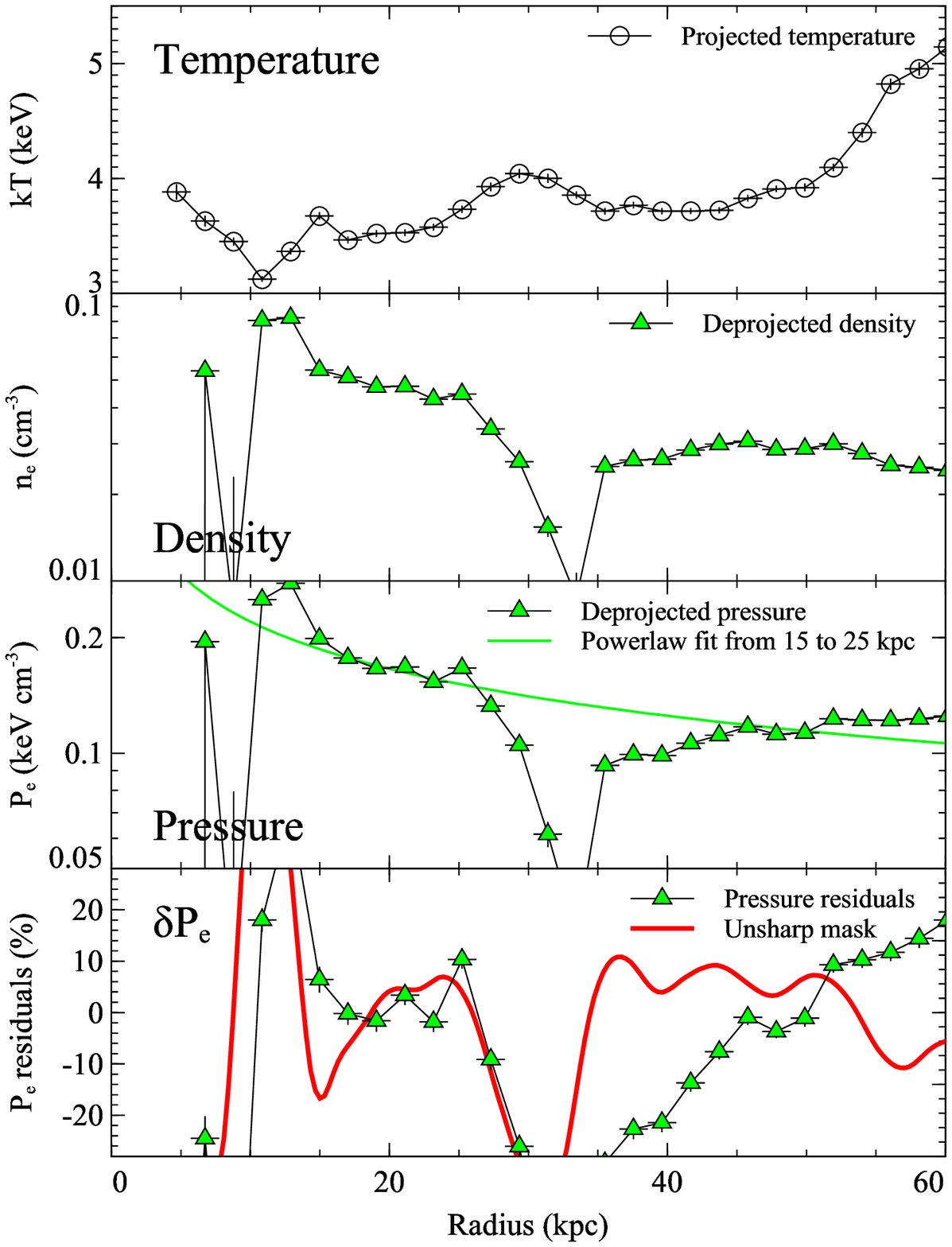}
  \caption{Temperature, density, pressure and pressure variation
    profiles. The red line shows the ripples from the unsharp mask
    image. The top left figure shows the profile in the NE direction,
    top right shows E direction, bottom left shows S direction and
    bottom right shows NW direction. The dashed line in the NE
    profiles indicates the position of the shock front.}
\end{figure*}

We now discuss the detailed behaviour of the temperature, density and
pressure around the shock surrounding the inner bubbles and the
ripples. Sectors have been defined to the NE, E, S and NW of the
nucleus and spectra extracted from bins spaced 5.4~arcsec in radius
(Table~\ref{tab:sectors}).

\begin{table}
  \begin{tabular}{lllll}
    Name & Centre RA & Centre Dec & Start angle & Stop angle \\ \hline
    North-east & 03:19:48.11 & +41:30:41.22 & 22.5 & 52.9 \\
    East &  03:19:48.11 & +41:30:41.22 & 91 & 106 \\
    South & 03:19:45.92 & +41:30:19.58 & 136.9 & 164.2 \\
    North-west &  03:19:48.11 & +41:30:41.22 & 294.4 & 334.2 \\
  \end{tabular}
  \caption{Sectors used to generate temperature, density and pressure
    profiles. Angles are measured from North in the eastern
    direction. Coordinates shown are J2000.}
  \label{tab:sectors}
\end{table}

\begin{figure}
  \includegraphics[width=\columnwidth]{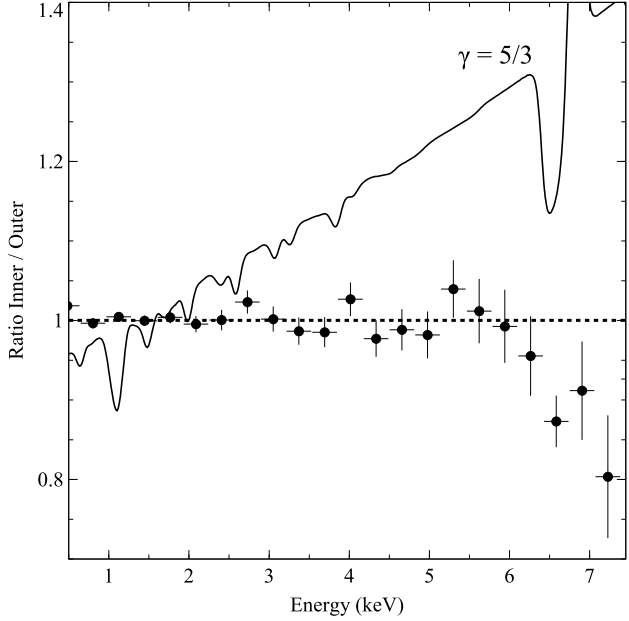}
  \includegraphics[width=\columnwidth]{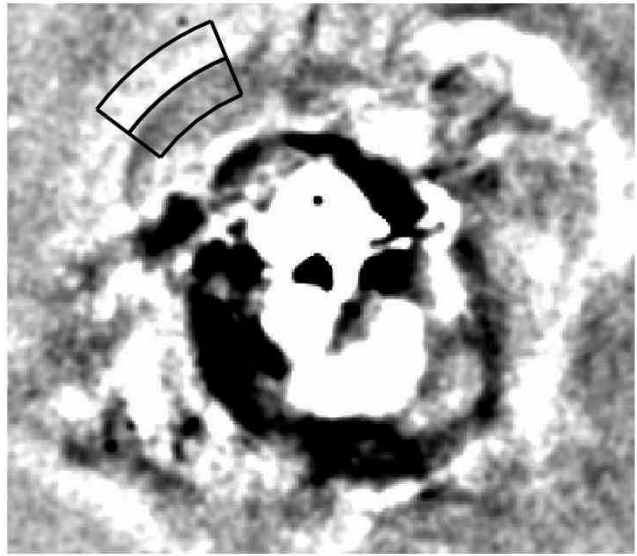}
  \caption{Top: The ratio of the spectrum inside the shock to that outside.
  The spectra have been normalised to have the same number of counts. The solid
line shows the ratio is the abrupt density jump corresponds to a weak
shock with $\gamma=5/3$. Bottom: Regions used in the above spectral
analysis, shown superimposed on an unsharp-masked image. The lower
region is that within the shock, the upper one is outside it. }
\end{figure}

\begin{figure}
  \includegraphics[width=\columnwidth]{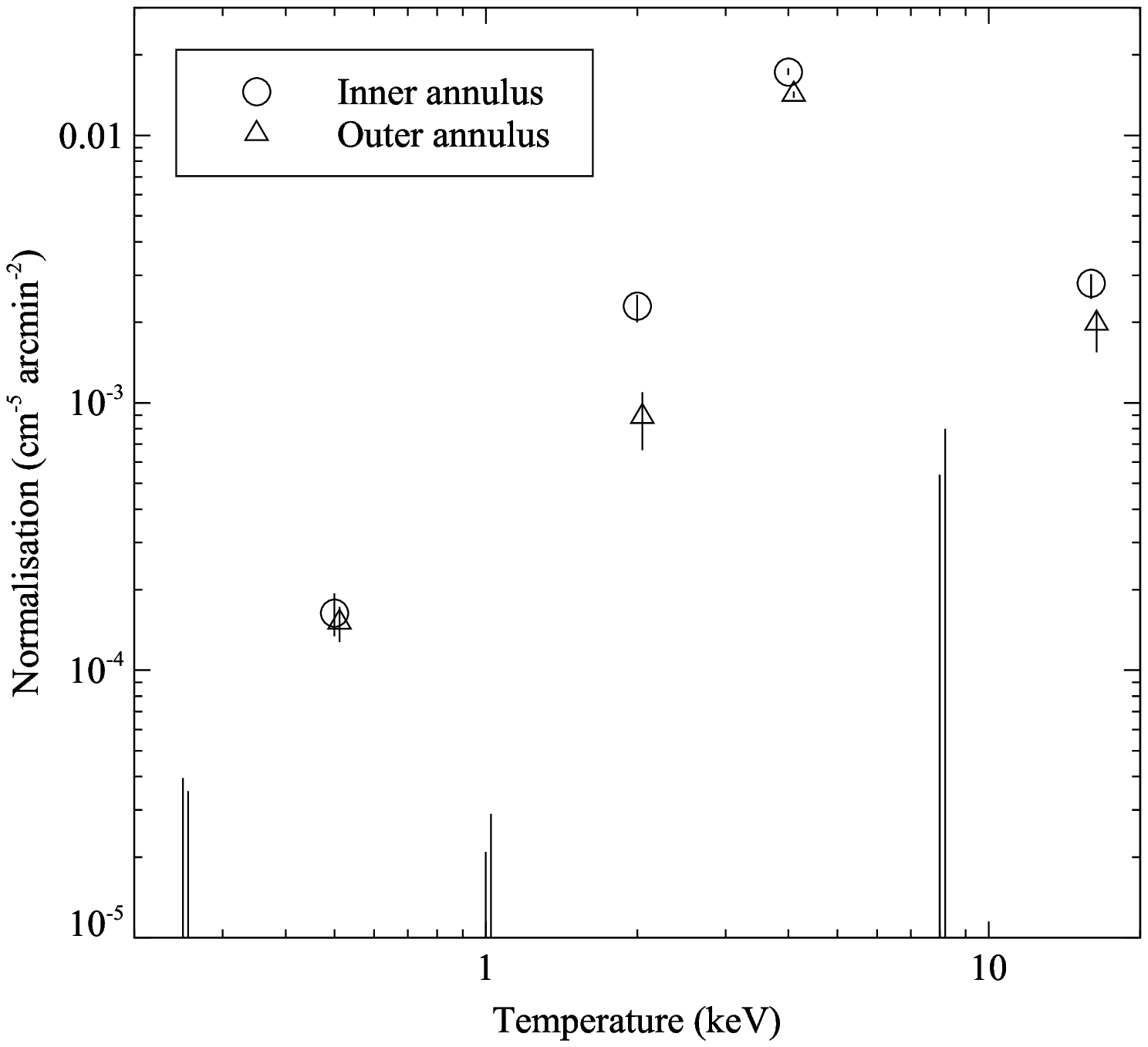}
  \caption{Distribution of temperature components within and outside
    the shock, using the same regions as shown in Fig.~9. The regions
have been fitted with models having temperature components at 0.25,
0.5, 1. 2. 4. 8 and 16~keV.
    Note that both the 4 and 2~keV components increase
    within the shock, with the 2~keV one increasing proportionately
    more. Only upper limits are obtained for the 0.25, 1 and 8~keV
components (the ends of the error bars are shown as the vertical lines).}
\end{figure}

The projected temperature profiles are shown in the top panels in
Fig.~8. We see generally that, apart from where the inner and ghost
bubbles lie, the temperature profiles are smoothly increasing from the
inside out.  We have explored deprojected temperatures using the
\textsc{xspec} \textsc{projct} routine but find them unstable with the
temperatures depending on which bins are used.  Dropping bins from
either the outside or the inside can have large unpredictable effects
on the results found for bins at intermediate radii. This may be due
to the real geometry being obviously different from the spherical or
ellipsoidal geometry assumed by the routine or to the gas being
multiphase. We therefore proceed to combine the projected temperatures
with approximately deprojected densities from the emission measures to
obtain pressure profiles. The deprojected densities are calculated by
subtracting the contribution to the fitted normalisation at each
radius by that expected from outer shells, assuming projection in a
spherical geometry.

The density and thus pressure profiles show variations (Fig.~8)
corresponding to the ripples seen in the unsharp mask image (Fig.~2).
The pressure residuals from a smooth power-law fitted from 20 to
70~kpc are shown in the lowest panel of the plots, together with the
residuals predicted from the unsharp mask image. Ripples in pressure
ranging from $\pm 5$ to $\pm 10$ per cent in pressure are seen out to
50~kpc or more. Such pressure variations cannot be static. They
resemble sound waves so, following our earlier work (Fabian et al
2003a), we interpret them as sound waves\footnote{Or at least they
operate as sound waves in a mixed thermal/relativistic plasma}
produced by the cyclic bubbling behaviour, or at least major expansion
episodes, of the inner bubbles. In the inner region they are high
pressure regions fronted by a weak shock, further out the shocks
weaken and are not distinguished from the overall pressure disturbance
or ripple.

A simple calculation serves to show the potential for the ripples to
heat the gas provided that viscosity is high enough to dissipate their
energy. Let us consider the region within 50~kpc where the ripples are
most clearly seen. If the ripples move at $1000\kmps$ then they cross
this region in $5\times 10^7\yr$. They cause the pressure to oscillate
with an amplitude of 5--10 per cent, which we conservatively take as 5
per cent of the thermal energy there. Consequently they can balance
cooling provided that the cooling time (roughly for the gas to lose
all its thermal energy) is 20 times the crossing time or a Gyr. This
condition is well met since the cooling time of the gas at 50~kpc is
2--3 Gyr dropping to about $2\times 10^8\yr$ near the centre (for the
hotter gas). The waves therefore need to dissipate about half their
energy by 50~kpc.  

The vertical dashed line in the upper panel of Fig.~8 corresponds to
the abrupt edge to the thick pressure band around the N bubble. It is
most clearly seen in this direction because there is little lower
energy emission superimposed upon it as appears to happen at other
azimuths. Our results show it to be associated with a jump in density
and this pressure and should be a shock front. However the projected
temperature hardly changes, even dropping slightly postshock (Fig.~8),
whereas it should rise on the basis of the factor 1.39 density jump,
from 4.04~keV to about 5.1~keV if the gas is adiabatic (see also
Fabian et al 2003a).

The very deep image enables us to now examine the exact temperature
structure across this shock front. Because of the geometrical
restrictions to any extended deprojection approach, mentioned above,
we compare spectra either side of the shock. All methods, including
single and two temperature, or using the outer emission as background
for the inner, show no sign of any hotter temperature
component within the shock in the range of 4--6~keV. The ratio of the
spectra post and preshock, plotted in Fig.~9, clearly shows that the emission
inside the shock is softer and inconsistent with the predicted rise,
if $\gamma=5/3.$ All the spectral fits have been carried out assuming
that the gas is in collisional and ionization equilibrium and
homogeneous within each region.

We have also carried out a multi-temperature analysis of the emission
both inside and outside the shock (Fig.~10). 56 per cent of the counts
per unit area across the jump going from out to inside is in an
increased 4~keV component and 28 per cent in a 2 keV one. There is
good evidence that the gas is multiphase. The 4~keV component, which
is the volume-filling one, shows an increase of density across the
shock but no increase in temperature.

The soft X-ray emission from the outer radial, N optical H$\alpha$
filament stops abruptly at the shock front. It is possible that the
2~keV component arises from shock heating of this cooler (1~keV)
gas. The pressure jump at the shock front indicates that this is a
weak shock so most of the heating is just $P \: \mathrm{d}V$ compression.

Why the 4~keV component does not change is puzzling if the gas is
adiabatic; much of the heating in a weak shock is just compressional
$P \: \mathrm{d}V$ heating. 

A fair approximation for the temperature behaviour expected in a shock
where the density jumps by the observed factor of 1.39 (Fig.~8) from
4.04~keV is that the postshock temperature $T=2.5 + 1.56\gamma\keV$. 
The observed behaviour of the gas is therefore explained if the 4~keV
component is isothermal with $\gamma\sim1$. 

A simple explanation for this would be that thermal conduction is
operating on this volume-filling gas phase.  The electrons, moving
faster than the ions, can go ahead of the shock (see e.g. Zel'dovich \&
Raizer 1966; Borkowski, Shull \& McKee 1989). If the magnetic field in
this region is mostly radial, then conduction can eliminate
temperature differences on a timescale 
\begin{equation} 
t_{\rm cond}\approx n k \ell^2/\kappa = 2.3\times 10^6 n\ell^2 T^{-5/2} \yr
\end{equation} which compares with the timescale for distance $\ell$
of matter to accumulate behind the shock (in the rest frame of that
matter) 
\begin{equation} 
t_{\rm shock}=4.8\times10^6 \ell \yr
\end{equation} 
where $\ell$ is the length inside the shock in units of
3~kpc (the bins in Figs.~8 and 9 are 2~kpc apart). The full Spitzer
(1956) rate for conduction is assumed here and the total postshock
density is in units of the observed value of $1\pcmcu$ (Fig.~8), the
postshock temperature (assuming an adiabatic gas) is in units of the
expected one of $T=5.1\keV= 6.3\times 10^7\K$. Magnetically-isolated blobs 
such as may comprise the lower temperature
component are compressed adiabatically. The post-shock timescale for
electron-ion equilibration is comparable to the above time ($\sim
2\times 10^6\yr$). This enhances the effect of conduction (halving it)
since only the electron energy needs to be conducted ahead of the
shock. We envisage that the
ion temperature jumps at the shock front but that the electron
temperature varies smoothly through this region, with a hotter
precursor extending  into the unshocked gas (see Fig.~7.19 on p519 of
Zel'dovich \& Raizer 1966); both the electron and
ion densities jump at the shock front.

This result introduces the possibility that thermal conduction is
effective in parts of the innermost regions of clusters. It has been
proposed and tested as a means for heating the gas from the outside,
but found to be inadequate for clusters and regions below 5~keV (Kim
\& Narayan 2003; Voigt \& Fabian 2004). What is needed in the Perseus
cluster is for thermal conduction to operate throughout much of the
inner hotter volume-filling phase. The ripples would therefore be {\it
  isothermal} sound waves (see Fabian et al 2005 for a comment on this
possibility). Both sound waves and conduction are then effective in
distributing the $P \: \mathrm{d}V$ work done by the bubbles into the
surrounding gas. Repeated bubbling in the central region may have
ordered the magnetic field into a roughly radial structure.

Cooler and/or cooling temperature components embedded in the hotter
gas behind the shock can damp any temperature rise behind the shock if
they mix with the hotter gas. The mass fraction of cooler gas required
(approx 30 per cent) appears not  to be high enough (see Fig.~10) for
this process to be important. It remains possible that mixing takes
place with larger masses of unseen cold gas which radiates much of the
thermal energy in yet unseen bands.

An issue which could be very important for shock propagation in the
inner intracluster medium is the presence of a relativistic plasma
(cosmic rays and magnetic field) in the inner core of the Perseus
cluster. This is evident here from the synchrotron emission seen as
the radio `mini-halo' (Pedlar et al 1990; Gitti et al 2003) and the
inverse Compton emission seen as a hard X-ray flux component (Sanders
et al 2005; it appears as the 16~keV component in Fig.~10). In the
collisionless conditions relevant to the shock it may be possible that
the relativistic plasma soaks up the energy, leaving the gas
isothermal. Indeed it could be repeated shocks from the bubbles which
reaccelerates the relativistic particles. They could redistribute the
energy to larger radii, serving to transport some of the energy and
creating a distributed heat source for the gas. We note that the
electron temperature observed behind the strong shocks in young
supernova remnants do not always fit expectations for simple
hydrodynamical shocks, probably due to particle acceleration (Rakowski
2005 and references therein). Although promising as a mechanism, there
are many uncertainties as to how it could operate and why there is no
sharp rise in either the synchrotron emission seen in radio maps nor
in the inverse Compton emission at the position of the shock. Moreover
it does not explain how the electrons avoid compressional heating.

The isothermal nature of the inner gas  raises the possibility
that the bubbles expand much faster than previously suspected from
observations. Initial models for the action of a central radio source
on the ICM by Heinz et al (1998) predicted that the bubbles would be
surrounded by shocks but \emph{Chandra} showed no evidence for shock-heated
gas. Efficient thermal conduction will however eliminate shock heating
as a diagnostic. Consequently the bubbles may expand, at times, faster
than inferred, even supersonically. The likely behaviour given the
variability of radio sources is that they expand in fits and starts,
with each rapid expansion phase giving rise to a sound wave. The
observation from the Perseus cluster of only one set of ghost bubbles
within 50~kpc radius yet 3 or more ripples allows for each bubbles to
generate several ripples before becoming buoyant enough to separate
and rise. This means that any estimate of bubbling power based simply
on buoyancy times (e.g. Birzan et al 2004; Dunn et al 2004, 2005a) is a
lower limit.

A further issue with regard to the energy injected by the bubbles is
the thickness of the postshock gas. This is very similar to the radius
of the bubbles, so has a volume about 7 times that of the bubbles
themselves. The pressure in the shocked gas is 30 per cent above the
outer unshocked gas, $P$, (Fig.~8) so the energy content of the
postshock gas is more than twice that obtained by assuming it is just
$PV$ where $V$ is the volume of a bubble. The work done ($\int
P\:\mathrm{d}V$) will be yet higher if some has been transported away
by conduction or relativistic particles.

\section{The multiphase nature of the gas}
Figs.~1 and 3 clearly show filamentary soft X-ray emission which is
closely associated with the optical H$\alpha$ filaments (Fabian et al
2003b). This soft emission has a temperature of between 0.5 and 1~keV
and would appear much brighter if the Galactic column density to the
Perseus cluster were not as high as the observed value of $\sim
1.3\times 10^{21}\pcmsq$. The Doppler velocities determined for the
filaments are $100-200\kmps$ and coherent over many kpc (Hatch et al
2005b) so, given their large radial extent and likely origin as being
pulled out from the centre by rising bubbles (Fabian et al 2003b), the
lifetimes of the filaments are several tens million yr, or even longer. 
In order to survive in the surrounding hot gas they must be insulated
from it or thermal evaporation would have caused them to disappear
within a million yr (equation 1). Conduction must therefore be highly
suppressed, by at least a factor of 100, probably due to magnetic
fields along their length (conduction is suppressed perpendicular to
the field direction).

As already mentioned, the filaments stop at the shock which is
probably disrupting them there. The filaments coincident with the
shock to the SE are probably just projected in front of the shock and
are not within it. Such magnetically isolated regions need not
completely vanish once they  are disrupted and may survive as higher
temperature blobs maintaining their isolation. The gas can therefore
be multiphase, not due to a thermal cooling instability, but to the
forced mixing of different components. Whether there is then slow
conductive evaporation or radiative condensation (see e.g. B\"ohringer
\& Fabian 1989) or turbulent mixing (e.g. Begelman \& Fabian 1990;
Loewenstein \& Fabian 1990) remains to be seen. 

\begin{figure}
  \includegraphics[width=\columnwidth]{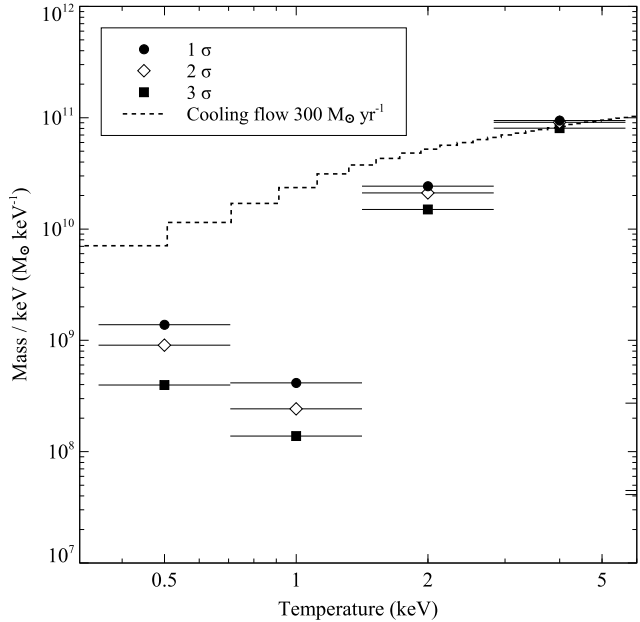}
  \caption{Distribution of mass at the fixed temperatures from within
    the innermost 1.5~arcmin radius. For comparison the expected
    result from a constant pressure cooling flow is shown.}
\end{figure}

\begin{figure*}
  \includegraphics[width=2\columnwidth]{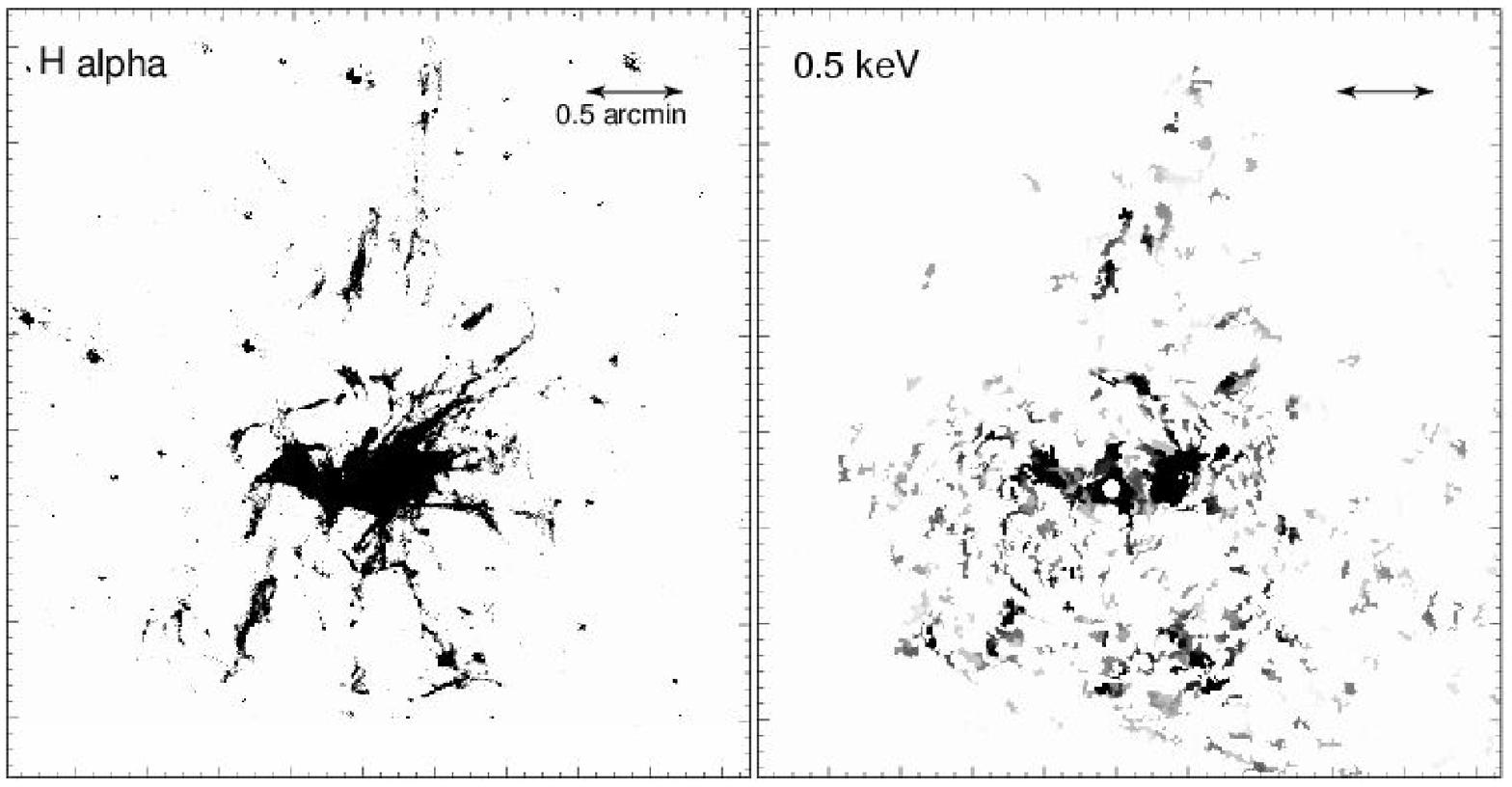}
  \includegraphics[width=2\columnwidth]{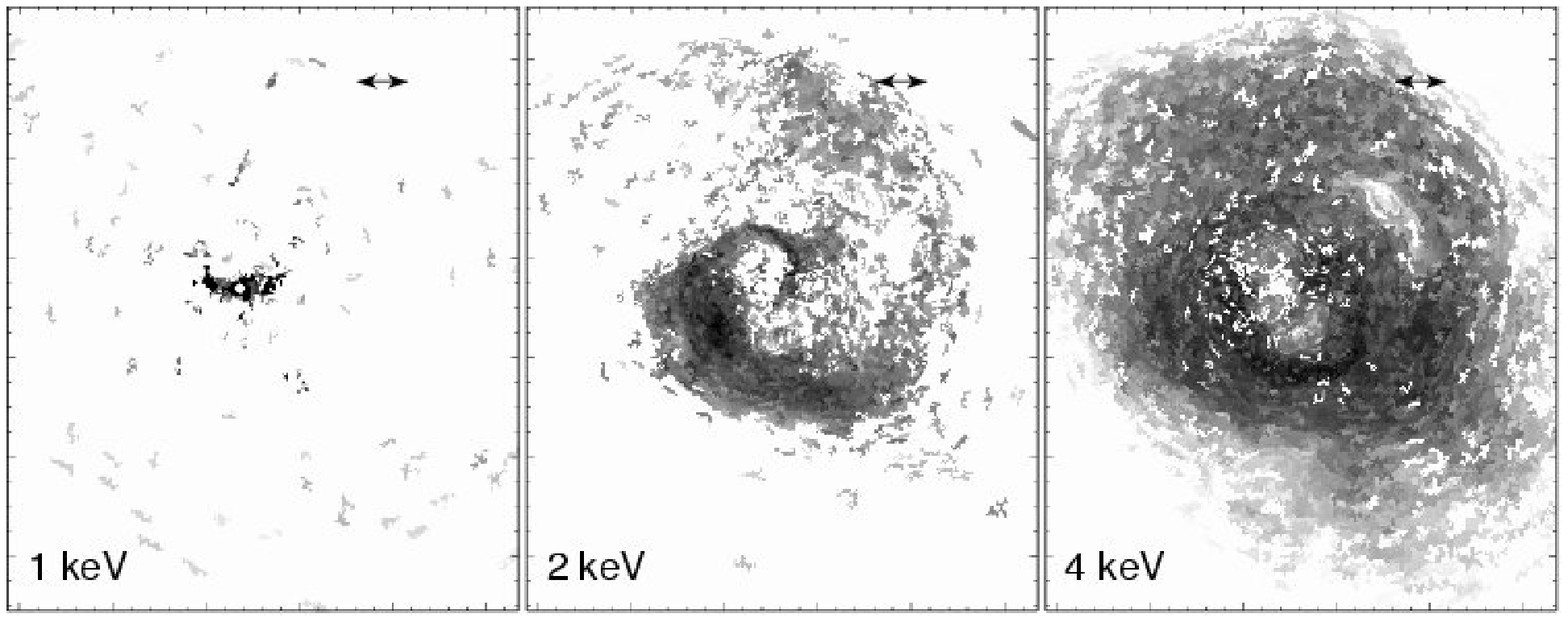}
  \caption{The H$\alpha$ filaments from the optical narrow-band image
    of Conselice et al (2000) is shown for comparison with maps of
mass in the 
    0.5, 1, 2 and 4~keV components. }
\end{figure*}

We have therefore conducted a multi-temperature determination of the
gas distribution in the Perseus core. The individual spectra generated
from regions chosen to contain $10^4$ counts or greater have been
fitted with a multi-temperature model consisting of gas at 0.5, 1, 2,
4, 8 and 16~keV (see also Sanders et al 2004 for similar fits to the
200~ks data).  The results have been mapped in terms of mass,
determined from the emission measure $n^2V$ divided by the density $n$
relevant for the pressure at that radius and measure temperature of
the component. Of course the volume filling factor of the gas
significantly different from the mean temperature found from the
single temperature fits (Fig.~4) is small. The widely differing mass
distributions on the Sky show that the gas is genuinely multiphase
(i.e. having different temperatures at the same radius) and we are not
mapping a mere projection effect.

We see a striking similarity in the 0.5~keV map to the optical
filaments with a total mass in the hot gas much larger than typically
found from an estimate of $3\times 10^7\Msun$ based on the total
luminosity of $H\alpha$ (Heckman et al 1989), a temperature of 5000\K\
for the gas and the surrounding pressure found here for the outer
filaments. The continuing pressure rise to the centre will reduce this
estimate and addition of molecular hydrogen, seen in the infrared
(e.g. Hatch et al 2005b and references therein) will tend to increase
it, so it should be a reasonable estimate. The mass maps at the
various temperatures are plotted in Fig.~12 and the total mass
distribution, determined from the masses within 1.5~arcmin of the
nucleus, is shown in Fig.~11. The different points at each temperature
show the total mass including only those which are
significant to 1, 2 and 3-$\sigma$. Noise will be a strong contaminant
to the lowest significance point.  For comparison, the mass
distribution expected for a steady cooling flow of $300\Msunpyr$ is
superimposed.

Interestingly, we find that there is a large drop-off in mass at 1~keV
but a recovery at around 0.5~keV. This rise is of course due to the
filament region. Until we know the fate of such material, in terms of
whether it is being heated or cooled by radiation or mixing, we cannot
say whether the bulk of the cooler gas, which lies in an E-W extended
clump around the nucleus, is the residual of a cooling flow or not.
We note that Bregman et al (2005) find OVI emission (characteristic of
gas at $5\times 10^5\K$) in a 30 arcsec \emph{FUSE} aperture centred
on the nucleus consistent with a mass cooling rate of about
$50\Msunpyr$.  This is comparable to the rate inferred from our mass
determination from gas at 0.5~keV (i.e. $\sim 5\times 10^6\K$), since
the radiative cooling time of gas between 0.5 and 2~keV in the inner
parts of the cluster is about $10^8 T^2\yr$, where $T$ is in keV.
Peterson (2005; private communication) finds a limit of only
$20\Msunpyr$ from a search for FeXVII emission in \emph{XMM-Newton}
RGS spectra of the inner 30 arcsec radius.

The fact that we see less gas at 1~keV could be the consequence of
cooling due to mixing, rather than radiation, dominating in that
temperature range. Such a possibility has been discussed by Fabian et
al (2002) and Soker, Blanton \& Sarazin (2002). The energy of the
hotter gas could in part go to heating the cooler gas at $\sim 10^4\K$
where there has long been a heating and excitation problem 
(Johnstone, Fabian \& Nulsen 1987; Heckman et al 1989; Sabra et al 2002). 
Indeed a mixing solution similar to a turbulent, radiative mixing layer
seems inevitable given the much lower mass in cold gas below $10^4\K$
then at 0.5~keV. 

A final inter-relationship between the hotter X-ray emitting gas and
the optical filaments is shown in Fig.~13. There is a partial ring
structure to the SE in the temperature swirl, resembling a letter `C'
written backwards. It coincides with some bright optical filaments and
in particularly with the `blue loop', first remarked on by Sandage
(1972) and seen well in many recent images (e.g. the blue band Jacobus
Kapteyn Telescope, JKT, image of Fig.~12). We presume that gas in the swirl
at this location collapsed and formed the stars in the astronomically
recent past.

\begin{figure}
  \includegraphics[width=\columnwidth]{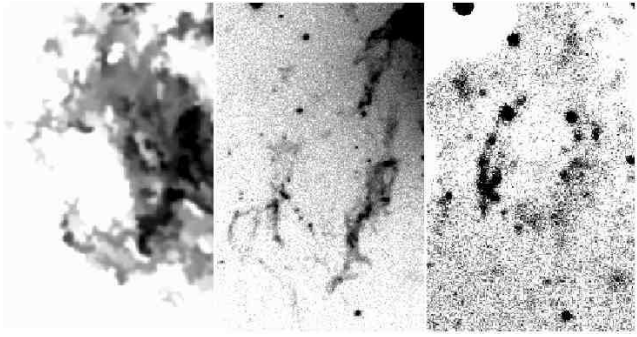}
  \caption{The blue loop is shown in X-ray temperature (left),
    H$\alpha$ (from Conselice et al 2001, centre) and blue light (from
    JKT; right).}
\end{figure}

Heinz \& Churazov (2005) have proposed that the relativistic component
discussed in Section 4 could exist in small blobs which could help to
dissipate wound waves. We see no obvious signs of small holes in the
X-ray emission larger than a few 100~pc in size. How well the
relativistic and thermal components are mixed is of importance for
transport processes in the region.

\section{Discussion}
We have found that the shock seen in our 200~ks image is isothermal.
The ripples seen beyond the shock are therefore likely to be
isothermal waves. Their energy is then dissipated by viscosity.
Conduction and sound waves can act together to dissipate and
distribute the energy from the radio source, and ultimately the
central massive accreting black hole. An isothermal shock allows
energetic bubbling to occur at the centre without overheating the
innermost region, a problem noted by Fujita \& Suzaki (2005) and
Mathews et al (2005).

In the work of Fujita \& Suzaki (2005) it is assumed that all wave
dissipation occurs at the shock front and does not include any later
dissipation via viscosity as the (observed) waves propagate further.
In one model they include conduction at 20 per cent of the Spitzer
rate and find agreement with the shape of the temperature and density
profile. However most of the energy in their model is supplied by
thermal conduction from the hotter outer gas; the AGN only dominates
over the region from 20--30~kpc where the shock occur. As they remark,
a double heating model with the AGN heating the inner regions and
conduction the outer was proposed earlier by Ruszkowski \& Begelman
(2002). 

It is clear from the temperature profile shown in the top left panel
of Fig.~8 that conduction of heat from the outer hotter gas is not
important within the inner 60~kpc in the Perseus cluster, since the
temperature profile is so flat. Indeed from 40--55~kpc the gradient
acts in exactly the wrong direction. As discussed in Section 4, the
observed ripples (which are strong sound waves or weak shocks) have
more than sufficient energy to heat the inner 50~kpc and so it is not
clear that any thermal conduction of heat from the outer gas is required.
What our analysis has shown is that thermal conduction acting in the
inner regions can account for the observed isothermal nature of the
shock and so prevent the problem of an accumulation of hot shocked
gas. The conduction merely acts to mediate the shock and redistribute
the energy from the central AGN.

The magnetic configuration of the field in the core is crucial to the
conductive behaviour.  We require an approximately radial field across
the shock, which is not understood. One possibility is that it arises
as a consequence of cooling and compression of the inner gas in the
past which leads to the frozen-in magnetic field being
predominately radial (Soker \& Sarazin 1990).  Nearby we have low
temperature H$\alpha$ emitting filaments which must be many 10s
million yr old and so magnetically isolated. We also find evidence for
multi-temperature, presumably multiphase, gas. The magnetic
connectedness is crucial to how the gas behaves. It raises the
possibility that the swirl seen in the temperature and entropy maps is
magnetically separate from the rest of the gas. Perhaps it is a fossil
from the merger of a galaxy with NGC\,1275 where the incoming (less
massive) galaxy `combed' the field into the apparent swirl. The 2 keV
gas immediately around the rim of the N inner bubble is presumably
protected from evaporation by a tangential field there.  Such
speculation may eventually be testable when it is possible with the
Expanded Very Large Array to carry out Faraday Rotation studies in
this region at higher frequencies and greater sensitivity than
currently feasible.  Preliminary indications from high resolution
studies of the nucleus with the Very Long Baseline Array indicate
fairly extreme Rotation Measures of up to ~7000 radian m$^{2}$ (Taylor
et al. 2005, in preparation).

We have also found a roughly N-S channel in the pressure difference
map which demonstrates the passage of a sequence of radio bubbles. The
outer ones are large and could be where they accumulate or just
represent a past, more energetic, period of activity. We also see part
of an unusual cold front to the S. This region is seen clearly in the
unsharp mask images (Fig.~2) and in one generated from data from all
chips (Fig.~13). This structure appears to be connected to a region to
the SW of the nucleus where the channel appears. It could represent
gas associated with subcluster merging in the cluster. Most likely
given the relationship with the bubble channel, the gas could be
evidence of past energetic bubbles. The bubble channel is good
evidence that the bubbles are not easily disrupted, presumably due to
the magnetic structure (De Young et al 2003) and/or viscosity in the
surrounding gas (Reynolds et al 2005). We assume that the pressure
dips in the channel because there is unseen buoyant relativistic plasma there
from the radio outbursts.

An overall picture of the region is shown in the image of Fig.~14,
where data from all chips has been used. The structure of the inner
regions can be seen together with the outer S bay, embedded within the
more extended peak of cluster X-ray emission. The upper part of the
H$\alpha$ fountain (Fig.~2) can also be more clearly seen.

\section{Summary}
Using a very deep, 900~ks \emph{Chandra} image of the core of the Perseus
cluster, we have found new outer features 50--80~kpc from the nucleus
and measured the detailed properties of gas near the centre. The
features are in the form of a concave cold front and baylike region of hot
gas which is in approximate pressure equilibrium. This could be the
result of an energetic past outburst from the nucleus, or where 
bubbles accumulate.

The inner radio bubbles are surrounded by complete higher pressure bands of
gas behind a sharp front. The gas temperature does not change across
the shock front, probably indicating that thermal conduction operates
efficiently there, or that co-existing relativistic plasma mediates
the shock. Pressure variations coincident with ripples
previously found in unsharp mask images reveal the presence of
isothermal sound waves. The isothermal nature of the innermost gas
means that a simple temperature estimate there does not reveal the
expansion velocity of the bubbles. We suspect that they expand in
rapid steps associated with outbursts of activity from the central
radio source. Provided that the energy in the ripples is dissipated by
viscosity, then the present heating rate in the ripples is sufficient
to balance radiative cooling. Larger pressure variations are seen
along a N-S channel, suggesting a sequence of bubbles, revealing the
activity of the central radio source for the past $10^8\yr$.

The gas in the centre is significantly multiphase with a large mass of
gas ($\sim 10^9\Msun$) associated with the optical H$\alpha$
filamentary nebula, with ten times more mass in 0.5~keV gas than that
radiating the optical emission lines. Mixing is likely occurring
between the hot ICM and the cold filamentary gas, with much radiative
cooling probably taking place below $10^6\K$. 

Cluster cores are complicated with the behaviour dependent on the
bubbling of a central radio source and on microphysical transport
processes. These in turn depend on the magnetic field structure, which
itself may be a consequence of past cooling and bubbling. 

\begin{figure*}
  \includegraphics[width=2\columnwidth]{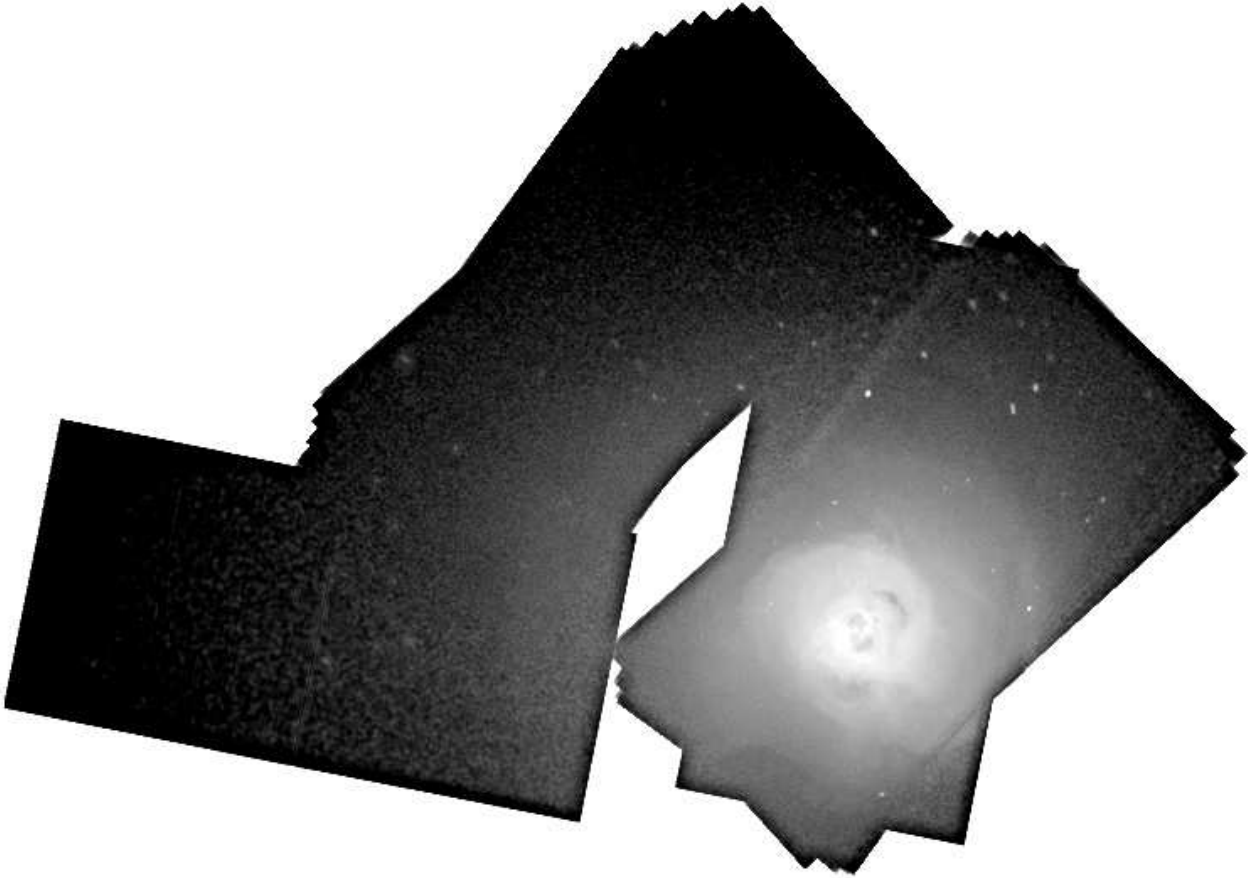}
  \caption{Total 0.5--7~keV image from all \emph{Chandra} CCD chips. }
\end{figure*}

\section{Acknowledgements} We thank the referee (Y. Fujita) for
comments. CSC and ACF thank the Royal Society for support. GBT
acknowledges support for this work from the National Aeronautics and
Space Administration through Chandra Award Number GO4-5134X issued by
the Chandra X-ray Observatory Center, which is operated by the
Smithsonian Astrophysical Observatory on behalf of the National
Aeronautics and Space Administration under contract NAS8-03060.  The
work of SWA is supported in part by the U.S. Department of Energy
under contract number DE-AC02-76SF00515.

\clearpage


\begin{thebibliography}{}
\bibitem[]{arnaud96} Arnaud K.A., 1996, Astronomical Data Analysis
  Software and Systems V, eds. Jacoby G. and Barnes J., p17, ASP Conf.
  Series volume 101 
\bibitem[]{} Anders E., Grevesse N., 1989, Geochimica et Cosmochimica Acta 53, 197
\bibitem[]{} Begelman M.C., Fabian A.C., 1990, MNRAS, 244 26P
\bibitem[]{} Birzan, L., Rafferty, D. A., McNamara, B. R., Wise, M. W.,
  Nulsen P. E. J. 2004, ApJ 607, 800
\bibitem[]{} B\"ohringer H, Fabian AC, 1989, MNRAS, 237, 1147
\bibitem[]{} B\"ohringer H., Voges, W., Fabian, A. C., Edge, A. C. and Neumann, D. M., 1993, MNRAS, 264, 25
\bibitem[]{} Borkowski KJ, Shull, M, McKee CF., 1989, ApJ, 336, 979 
\bibitem[]{} Bregman, J.D., Fabian A.C., Miller D.D., Irwin J.A.,
2005, ApJ submitted 
\bibitem[]{} Churazov E., Forman, W., Jones, C., and B\" ohringer, H., 2000, A\&A, 356, 788
\bibitem[]{} Conselice C.J., Gallagher J.S., Wyse R.F.G., 2001, ApJ,
559, 791
\bibitem[]{} Dunn RJH, Fabian, A. C., 2004, MNRAS, 355, 862
\bibitem[]{} Dunn RJH, Fabian, A. C., Taylor GB., 2005a, MNRAS in press
(astro-ph/0510191)
\bibitem[]{} Dunn RJH, Fabian, A. C., Sanders JS., 2005b, MNRAS submitted
\bibitem []{} Fabian A.C., Celotti A., Blundell K.M., Kassim N.E.,
Perley R.A., 2002, MNRAS, 331, 369
\bibitem[]{} Fabian AC., Sanders, J.S., Allen, S.W.,  Crawford, C.S., Iwasawa, K., Johnstone, R.M., Schmidt, R.W., and Taylor, G.B., 2003(a), MNRAS, 344L, 43F
\bibitem[]{} Fabian AC., Sanders, J.S., Crawford, C.S., Conselice,  C.J., Gallagher, J.S., and Wyse, R.F.G., 2003(b), MNRAS, 344L, 48
\bibitem[]{}  Fabian, A. C., Sanders, J. S., Ettori, S., Taylor, G. B., Allen,
S. W., Crawford, C. S., Iwasawa, K., Johnstone, R. M. and Ogle, P. M.,
2000, MNRAS 318, L65
\bibitem[]{} Fabian AC., Reynolds CS., Taylor GB., Dunn RJH.,  2005,
MNRAS, 363, 891
\bibitem[]{} Fabian AC, Allen SW, Crawford CS, Johnstone RM, MOrris
GM, Sanders JS, Schmidt RG.,  2002, MNRAS, 332, L50
\bibitem[]{} Forman W. et al 2003, astro-ph/0312576
\bibitem[]{} Fujita Y., Suzuki TK., 2005, ApJ, 630, L1
\bibitem[]{} Gillmon K., Sanders J.S., Fabian A.C., 2004, MNRAS, 348,
159 
\bibitem[]{} Gitti M, Brunetti G, Setti G., 2002, A\&A, 386, 456
\bibitem[]{} Hatch NA, Crawford CS, Fabian AC, Johnstone RM., 2005a,
MNRAS, 358, 765
\bibitem[]{} Hatch NA., Crawford CS, Johnstone RM., Fabian
AC., 2005b, MNRAS, submitted 
\bibitem[]{} Heckman TM., Baum SA., van Breugel WJH., McCarthy P.,
1989, ApJ, 338, 48 
\bibitem[]{} Heinz S., Reynolds CS., Begelman MC., ApJ, 1998,
501, 126
\bibitem[]{} Heinz S, Churazov E, 2005 astro-ph/0507038 
\bibitem[]{} Johnstone R.M., Fabian A.C., Nulsen P.E.J., 1987, MNRAS, 224, 75 
\bibitem[]{} Kim W-T., Narayan R., 2003, ApJ, 596, L139
\bibitem{liedahl95} Liedahl D.A., Osterheld A.L., Goldstein W.H.,
  1995, ApJ, 438, L115
\bibitem[]{} Loewenstein M., Fabian AC., 1990, MNRAS, 242, 120
\bibitem[]{} Mathews, W.G., Faltenbacher, A., Brighenti, F., 2005, ApJ,
in press (astro-ph/0511151)
\bibitem[]{} Markevitch M., et al 2000, ApJ, 541, 542
\bibitem[]{} McNamara B R., Nulsen, P. E. J., Wise, M. W., Rafferty, D. A.,
Carilli, C., Sarazin, C. L. and Blanton, E. L., 2005, Nature, 433, 45
\bibitem{mewe85} Mewe R., Gronenschild E.H.B.M., van den Oord
  G.H.J., 1985, A\&AS, 62, 197
\bibitem[]{} Pedlar A., Ghataure HS., Davies RD., Harrison BA., Perley
R., Crane PC., Unger SW., 1990, MNRAS, 246, 477
\bibitem[]{} Rakowski C., 2005, Adv Space Res., 35, 1017
\bibitem[]{} Reynolds., McKernan, B., Fabian, A.C., Stone J.M., and Vernaleo,J.C., 2004, MNRAS, 
\bibitem []{} Ruszkowski M., Begelman M.C., 2002, ApJ, 581, 223
\bibitem []{} Ruszkowski M., Br\"uggen M., Begelman M.C., 2004a, ApJ,
  611,158
\bibitem []{} Ruszkowski M., Br\"uggen M., Begelman M.C., 2004b, ApJ,
  615,675
\bibitem []{} Ruszkowski M., Br\"uggen M., Hallman E.,, 2005, astro-ph/0501175
\bibitem[]{} Sabra BM., Shields JC., Filippenko AV., 2000, ApJ, 545,
157
\bibitem[]{} Sandage A.R., 1972, in Nuclei of Galaxies, ed O'Connell
\bibitem[]{} Sanders J.S., Fabian A.C., Allen S.W., Schmidt R.W.,
  2004, MNRAS, 349, 952
\bibitem[]{} Sanders J.S., Fabian A.C., Dunn R.J.H., 2005, MNRAS, 360,
  133
\bibitem[]{} Schmidt R W., Fabian AC., Sanders JS., 2002, MNRAS, 337, 71
\bibitem[]{} Soker N., Blanton, E.L., Sarazin C.~L., 2004, A\&A, 422,
445
\bibitem[]{} Soker N., Sarazin C.~L., 1990, ApJ, 348, 73 
\bibitem[]{} Spitzer L., 1956, Physics of Fully Ionized Gases, 1st
  Ed., New York: Wiley-Interscience
\bibitem[]{townsley02a} Townsley L.K., Broos P.S., Chartas G.,
  Moskalenko E., Nousek J.A., Pavolv G.G., 2002, Nuc. Instr. and
  Meth. in Phys. Res. A, 486, 716
\bibitem[]{townsley02b} Townsley L.K., Broos P.S., Nousek J.A., Garmire
  G.P., 2002, Nuc. Instr. and Meth. in Phys. Res. A, 486, 751 
\bibitem[]{} Voigt LM Fabian AC., 2004, MNRAS, 347, 1130
\bibitem[]{} Zel'dovich Y.B., Raizer Y.P., 1966, Physics of Shock
Waves  and High-Temperature Hydrodynamic Phenomena, Academic Press,
New York

\end{thebibliography}
\end{document}